\documentclass[11pt]{article}

\usepackage{amsthm}
\usepackage{amsmath}
\usepackage{amsfonts}
\usepackage{mathrsfs}
\usepackage{dsfont}

\usepackage{amssymb}
\usepackage{enumerate}
\usepackage{graphicx}
\usepackage{mathrsfs}

\usepackage{a4}
\usepackage{url}
\usepackage{algorithm}
\usepackage{algorithmicx}
\usepackage{algpseudocode}
\usepackage{color}

\usepackage[authoryear]{natbib}

\newcommand{\bea}{\begin{eqnarray*}}
\newcommand{\eea}{\end{eqnarray*}}
\newcommand{\be}{\begin{eqnarray}}
\newcommand{\ee}{\end{eqnarray}}

\def\dd{\mathrm{d}}

\def\Arg{\mathrm{Arg}}

\def\tr{\mathrm{trace}}

\def\mt{\theta}

\DeclareMathOperator{\CR}{\mathsf{CR}}
\DeclareMathOperator{\PR}{\mathsf{PR}}

\def\IMSE{\mathsf{IMSE}}
\def\ISE{\mathsf{ISE}}
\def\hIMSE{\widehat{\mathsf{IMSE}}}
\def\hISE{\widehat{\mathsf{ISE}}}

\def\KH{\mathsf{KH}}
\def\MN{\mathsf{MN}}
\def\Ex{\mathsf{E}}

\def\bDelta{\overline{\Delta}}

\def\ma{\alpha}
\def\mg{\gamma}
\def\ml{\lambda}
\def\mt{\theta}
\def\mtb{\boldsymbol{\theta}}

\def\mve{\varepsilon}
\def\ms{\sigma}
\def\Ex{\mathsf{E}}

\def\ra{\rightarrow}
\def\TT{^\top}
\def\e1{\mathsf{e}}

\def\Db{\mathbf{D}}

\def\Ib{\mathbf{I}}

\def\Kb{\mathbf{K}}
\def\bK{\overline{K}}

\def\bmOb{\overline{\mathbf{\Omega}}}
\def\bmGb{\overline{\mathbf{\Gamma}}}

\def\bmob{\overline{\mathbf{\omega}}}

\def\Qb{\mathbf{Q}}
\def\Rb{\mathbf{R}}
\def\Sb{\mathbf{S}}
\def\Xb{\mathbf{X}}

\def\Zb{\mathbf{Z}}

\def\1b{\mathbf{1}}
\def\0b{\mathbf{0}}
\def\cb{\mathbf{c}}
\def\Cb{\mathbf{C}}

\def\eb{\mathbf{e}}

\def\hb{\mathbf{h}}
\def\kb{\mathbf{k}}
\def\pb{\mathbf{p}}
\def\ub{\mathbf{u}}
\def\vb{\mathbf{v}}
\def\wb{\mathbf{w}}
\def\xb{\mathbf{x}}
\def\yb{\mathbf{y}}
\def\zb{\mathbf{z}}
\def\betab{\boldsymbol{\beta}}

\def\SE{{\mathscr E}}
\def\SF{\mathcal{F}}
\def\SSF{\mathscr{F}}
\def\SH{\mathcal{H}}
\def\SSL{\mathds{L}}

\def\SN{{\mathscr N}}
\def\SO{\mathcal{O}}

\def\SS{{\mathscr S}}

\def\SX{{\mathscr X}}

\newcommand{\fin}{\mbox{}~\hfill\mbox{$\lhd$}}
\newcommand{\carre}{\mbox{}~\hfill\rule{2mm}{2mm}}

\newtheorem{thm}{Theorem}

\newcommand{\vsp}{\vspace{0.3cm}}

\begin{document}

\title{Validation design I:\\
construction of validation designs via kernel herding}

\author{Luc Pronzato \& Maria-Jo\~ao Rendas \\
\mbox{}\\
CNRS, Universit\'e C\^ote d'Azur, Laboratoire I3S\\
B\^at.\ Euclide, Les Algorithmes, 2000 route des lucioles,\\
06900 Sophia Antipolis cedex, France \\
{\tt \{luc.pronzato,rendas\}@univ-cotedazur.fr}}

\maketitle

\begin{abstract}
We construct validation designs $\Zb_m$ aimed at estimating the integrated squared prediction error of a given design $\Xb_n$. Our approach is based on the minimization of a maximum mean discrepancy for a particular kernel, conditional on $\Xb_n$, so that sequences of nested validation designs can be constructed incrementally by kernel herding. Numerical experiments show that key features for a good validation design are its space-filling properties, in order to fill the holes left by $\Xb_n$ and properly explore the whole design space, and the suitable weighting of its points, since evaluations far from $\Xb_n$ tend to overestimate the global error. A dedicated weighting method, based on a particular kernel, is proposed. Numerical simulations with random functions show the superiority the method over more traditional validation based on random designs, low-discrepancy sequences, or leave-one-out cross validation.
\end{abstract}

{\small {\bf keywords} validation; design of experiments; computer experiments; discrepancy; space-filling design; greedy algorithm }


\section{Introduction and motivation}

This paper proposes methods to define designs enabling good estimation of the prediction performance of a given non-parametric model, which has been adjusted to a known training dataset.
More precisely, we suppose that a design $\Xb_n=\{\xb_1,\ldots,\xb_n\}$ with $n$ points in $\SX=[0,1]^d$ has been been used to build a predictor of the value of an unknown function $f$ on $\SX$. We denote
by $\yb_n=[f(\xb_1),\ldots,f(\xb_n)]\TT$ the vector collecting the $n$ evaluations of $f$ at the $\xb_i$ and by $\eta_n(\xb)=\eta_{[\Xb_n,\yb_n]}(\xb)$ the corresponding prediction of $f(\xb)$.
The Integrated Squared Error (ISE) over $\SX$ is then
\be\label{ISE}
\ISE(\Xb_n) = \int_\SX \left[\eta_n(\xb)-f(\xb)\right]^2 \, \mu(\dd \xb) \,.
\ee
Above, the measure $\mu$ codes the user preferences, penalizing regions of $\SX$ which are of particular interest or importance. In the paper we will always consider that $\mu$ is the Lebesgue measure on $\SX$, the extension to non-uniform $\mu$ requiring only minor modifications. Note that we slightly abuse notation here, as the dependency of $\ISE(\Xb_n)$ on $\Xb_n$ is hidden in $\eta_n(\cdot)$, which is adapted to the training set $\Xb_n$. The same shortcut is used throughout the paper. 

In practice, the integral in definition \eqref{ISE} is approximated by a discrete sum, which is equivalent to letting $\mu$ be a discrete measure with finite support $\Zb_m=\{\zb_1,\ldots\zb_m\}\subset\SX$, at which $\eta_n$ and $f$ are effectively evaluated.
The objective of the paper is to propose methods for the construction of \emph{validation designs} $\Zb_m$ and investigate the properties of the corresponding estimates of $\ISE(\Xb_n)$\footnote{Here the integral \eqref{ISE} will be estimated directly by a discrete sum over $\Zb_m$. The situation is different when the validation design $\Zb_m$ is used to predict the behavior of the error process $\mve_n(\xb)=\eta_n(\xb)-f(\xb)$, in order to estimate $\ISE(\Xb_n)$ by $\int_\SX \widehat\mve_n^2(x) \, \mu(\dd \xb)$. This alternative construction will be considered in a companion paper.}.

If $f$ were known, we could compute both $\ISE(\Xb_n)$ and its finite approximation
\bea
\hISE(\Zb_m,\Xb_n) = \frac{1}{m} \sum_{i=1}^m \left[\eta_n(\zb_i)-f(\zb_i)\right]^2\,,
\eea
and directly choose $\Zb_m$ to have $\hISE(\Zb_m,\Xb_n) \approx \ISE(\Xb_n)$ (even if selecting such $m$ points $\Zb_m$ would not be an easy task). However, $f$ is unknown and both $\ISE(\Xb_n)$ and $\hISE(\Zb_m,\Xb_n)$ can only be estimated. To do that, we shall adopt the kriging framework, which we briefly recall in Section~\ref{S:kriging}.

In our study, we consider that the design $\Xb_n$ is given, making no assumption on how it has been chosen\footnote{Methods similar to those we propose for the construction of $\Zb_m$ can also be used to construct $\Xb_n$; see Section~\ref{S:kh-basics} and the examples in Section~\ref{S:Examples}.}. We are interested in particular in situations where $m$ is not specified in advance and one wishes to construct an increasing sequence of imbedded designs $\Zb_k\subset \Zb_{k+1} \subset \Zb_{k+2} \subset \cdots$ such that the $\Zb_k$ have increasingly good performance as estimators of $\ISE(\Xb_n)$ when $k$ increases. As a consequence of the
underlying Gaussian framework chosen, the methods proposed for the construction of $\Zb_m$ will not depend on the function evaluations $\yb_n$.

The paper is organized as follows. A criterion measuring the quality of a validation design, based on Gaussian process modelling, is introduced in Section~\ref{S:kriging}. In Section~\ref{S:kh} we see how the proposed criterion can be optimized by kernel herding, detailing application of the general algorithm to it. The properties of the proposed design construction are investigated numerically
in Section~\ref{S:Examples}, exposing two important features: the completed design $\Xb_m\cup\Zb_m$ must be space-filling, the contributions of the individual errors in $\hISE(\Zb_m,\Xb_n)$ must be under-weighted to avoid overestimation of $\ISE(\Xb_n)$. These findings are confirmed in Section~\ref{S:numerical-results} where random test functions are used to illustrate achieved validation performance: the estimation of $\ISE(\Xb_n)$ is significantly more accurate than with leave-one-out cross validation or uniformly weighted random or space-filling designs, which all seriously overestimate $\ISE(\Xb_n)$.

\section{An ISE-based criterion for validation design}\label{S:kriging}

\subsection{A Gaussian process model}\label{S:GP}

As mentioned above, $f$ is unknown and we cannot choose $\Zb_m$ by minimizing $|\hISE(\Zb_m,\Xb_n)-\ISE(\Xb_n)|$ directly. Assumptions on the behavior of $f$ must be made. Considering the worst case for $f$ in a given class of functions would be an option. Here we shall follow another, simpler, route and assume that $f$ is the realization of a Gaussian Process (GP), or Gaussian Random Field, $\SF_x$ indexed by $\SX$, with given second-order characteristics.

For the sake of simplicity, we suppose that $\Ex\{\SF_x\}=0$ for all $\xb\in\SX$. Extension to the case of a linearly parameterized mean, with $\Ex\{\SF_x\}=\betab\TT\hb(\xb)$ for a vector $\betab$ of unknown parameters and a vector $\hb(\xb)=[h_1(\xb),\ldots,h_p(\xb)]\TT$ of $p$ known functions of $\xb$ (including the constant) is possible via some adaptation. We also suppose that $\Ex\{\SF_x \SF_{x'}\}=K(\xb,\xb')$, a known covariance function.
In practice, $K$ may be known up to a (variance) scaling coefficient $\ms^2$ and parameterized by some parameters $\mtb$, setting in particular the correlation lengths
and the smoothness of the functions that belong to the Reproducing Hilbert Space (RKHS) $\SH_K$ associated with $K$. Both $\ms^2$ and $\mtb$ can be estimated from the data $\SSF_n=\{\Xb_n,\yb_n\}$, e.g.\ by maximum likelihood, see for instance \citet{SantnerWN2003}, or cross validation; see \cite{Bachoc2013} and Section~\ref{S:Numerical results} for an example.

The GP assumption defines a prior distribution for $f$, which can be updated given $\SSF_n$ into a posterior distribution, with mean $\Ex\{\SF_x|\SSF_n\}=\kb_n\TT(\xb)\Kb_n^{-1}\yb_n$ and covariance
\be \label{covn}
\Ex\{\SF_x \SF_{x'}|\SSF_n\}=K_{|n}(\xb,\xb')=K(\xb,\xb')-\kb_n\TT(\xb)\Kb_n^{-1}\kb_n(\xb')\geq 0 \,,
\ee
for any $\xb$, $\xb'$ in $\SX$,  where
\bea
\kb_n(\xb) &=& \left[K(\xb,\xb_1)\,\ldots,K(\xb,\xb_n)\right]\TT \,,\\
\{\Kb_n\}_{i,j} &=& K(\xb_i,\xb_j)\,, \ i,j=1,\ldots,n\,,
\eea
the $n\times n$ matrix $\Kb_n$ being positive definite. The Integrated Mean Squared Error (IMSE)
\bea
\int_\SX  \Ex\left\{\left[\eta_n(\xb)-f(\xb)\right]^2|\SSF_n\right\}\, \mu(\dd \xb) &=&
\int_\SX  \Ex\left\{\left[\eta_n(\xb)-\kb_n\TT(\xb)\Kb_n^{-1}\yb_n\right]^2|\SSF_n\right\}\, \mu(\dd \xb) \\
&& + \int_\SX K_{|n}(\xb,\xb) \, \mu(\dd \xb)
\eea
is minimum when the prediction $\eta_n(\xb)$ equals the posterior mean $\kb_n(\xb)\TT\Kb_n^{-1}\yb_n$, which yields
\be\label{IMSE}
\IMSE(\Xb_n) = \int_\SX K_{|n}(\xb,\xb) \, \mu(\dd \xb) \,.
\ee
Note that $K_{|n}(\xb,\xb_i)=0$ for any design point $\xb_i$ and any $\xb$ in $\SX$ and that $\eta_n$ interpolates the observations $\yb_n$. The extension to the case where $\eta_n$ is not a interpolator does not raise particular difficulties, only yielding a kernel $\bK_{|n}$ different from \eqref{bK_|n}, having a slightly more complicated expression where the errors $f(\xb_i)-\eta_m(\xb_i)$ intervene.

\subsection{Validation designs minimizing the expected squared $\ISE$ difference}\label{S:IMSE-based}

Replacing $f(\xb)$ by a realization $\SF_x$ of the GP model of Section~\ref{S:GP} in $\ISE(\Xb_n)$ and $\hISE(\Zb_m,\Xb_n)$, we define
\bea
\bDelta^2(\Zb_m,\Xb_n) &=& \Ex\left\{ \left[ \ISE(\Xb_n) - \hISE(\Zb_m,\Xb_n) \right]^2 |\SSF_n \right\} \\
&=& \Ex\left\{ \left[ \int_\SX [\SF_x-\eta_n(\xb)]^2\, (\zeta_m-\mu)(\dd\xb) \right]^2 |\SSF_n \right\} \,,
\eea
the mean squared error of the ISE estimator $\hISE(\Zb_m,\Xb_n)$, and where $\zeta_m$ denotes the discrete measure $\zeta_m=(1/m)\,\sum_{i=1}^m \delta_{\zb_i}$, with $\delta_\zb$ the delta measure at $\zb$
($\zeta_m-\mu$ is thus a signed measure with total mass 0).

For any positive definite kernel $C(\cdot,\cdot)$ and probability measures $\xi$ and $\nu$, denote by $\mg_C(\xi,\nu)$ the Maximum Mean Discrepancy (MMD) between $\xi$ and $\nu$, defined by
\bea
\mg_C^2(\xi,\nu)=\int_{\SX^2} C(\xb,\xb') \, (\xi-\nu)(\dd\xb)(\xi-\nu)(\dd\xb') \,;
\eea
see \cite[Def.~10]{SejdinovicSGF2013}. The minimization of $\mg_C(\xi,\nu)$ with respect to $\xi$ for a given $\nu$ can be performed by kernel herding (Section~\ref{S:kh}), yielding a sequence of finitely supported measures $\xi^{(t)}$ such that $\mg_C(\xi^{(t)},\nu)\ra 0$ as $t\ra\infty$.

When $\eta_n(\xb)=\kb_n(\xb)\TT\Kb_n^{-1}\yb_n$, direct calculation gives
\be
\bDelta^2(\Zb_m,\Xb_n) &=& \int_{\SX^2} \Ex\left\{ [\SF_x-\eta_n(\xb)]^2[\SF_{x'}-\eta_n(\xb')]^2 |\SSF_n \right\}\, (\zeta_m-\mu)(\dd\xb)(\zeta_m-\mu)(\dd\xb') \nonumber \\
&=& \int_{\SX^2} \bK_{|n}(\xb,\xb') \, (\zeta_m-\mu)(\dd\xb)(\zeta_m-\mu)(\dd\xb') = \mg_{\bK_{|n}}^2(\zeta_m,\mu) \label{delta-MMD}
\ee
where, for all $\xb,\xb'$ in $\SX$, we denote
\be
\bK_{|n}(\xb,\xb')  = 2\, K_{|n}^2(\xb,\xb') + K_{|n}(\xb,\xb)K_{|n}(\xb',\xb')\,, \label{bK_|n}
\ee
with $K_{|n}$ defined by \eqref{covn}.
Note that $\bK_{|n}$ is positive definite (but not strictly positive definite, see Appendix~A).
Indeed, the Hadamard product $\Cb_n^{\circ 2}$ with elements $\{\Cb_n^{\circ 2}\}_{i,j}=C^2(\xb_i,\xb_j)$, $i,j=1,\ldots,n$, is positive definite when the matrix $\Cb_n$ with elements $\{\Cb_n\}_{i,j}=C(\xb_i,\xb_j)$ is positive definite. Hence, $K_{|n}^2$ is positive definite since $K_{|n}$ is positive definite, which implies that $\bK_{|n}$ is positive definite.
The fact that $\bDelta^2(\Zb_m,\Xb_n)$ does not depend on $\yb_n$ although it relies on conditioning on $\SSF_n$ is a direct consequence of using a GP model.

%

\section{Kernel herding for validation designs}\label{S:kh}

\subsection{A summary of kernel herding}\label{S:kh-basics}

Let $C$ denote a positive definite kernel. For any signed measure $\xi$ on $\SX$, let
\be\label{energy}
\SE_C(\xi)=\int_{\SX^2} C(\xb,\xb')\,\xi(\dd\xb)\xi(\dd\xb') \geq 0
\ee
denote the energy of $\xi$ for $C$, so that $\mg_{\bK_{|n}}^2(\zeta_m,\mu)=\SE_{\bK_{|n}}(\zeta_m-\mu)$ in \eqref{delta-MMD}.
A kernel $C$ is called \emph{characteristic} when $\mg_C(\cdot,\cdot)$ defines a metric on the set of probability measures on $\SX$, implying in particular that, for two probability measures $\zeta$ and $\mu$, $\mg_C(\zeta,\mu)=0$ if and only if $\zeta=\mu$. The kernel $\bK_{|n}$ is not characteristic, see Appendix~A, but we can nevertheless consider the minimization of $\mg_{\bK_{|n}}^2(\zeta_m,\mu)$. 


For any $\ma\in[0,1]$, we have
$(1-\ma)\,\SE_C(\xi)+\ma\,\SE_C(\nu) - \SE_C[(1-\ma)\xi+\ma\nu] = \ma(1-\ma)\, \SE_C(\xi-\nu) \geq 0$, showing that $\SE_C(\cdot)$ is convex; see \cite{PZ2020-SIAM}, and we can minimize the squared MMD criterion $\mg_C^2(\xi,\mu)=\SE_C(\xi-\mu)$ with respect to $\xi$ by a simple descent algorithm.

Denote by $F_{C,\mu}(\xi;\nu)$ the directional derivative of $\mg_C^2(\cdot,\mu)$ at $\xi$ in the direction $\nu$,
\bea
F_{C,\mu}(\xi;\nu) = \lim_{\ma\ra 0^+} \frac{\SE_C[(1-\ma)\xi+\ma\nu-\mu]-\SE_C(\xi-\mu)}{\ma} \,.
\eea
Straightforward calculation gives
\bea
F_{C,\mu}(\xi;\nu) = 2 \left[ \int_{\SX^2} C(\xb,\xb')\, (\nu-\mu)(\dd\xb)(\xi-\mu)(\dd\xb') - \SE_C(\xi-\mu) \right] \,.
\eea
In particular, for $\nu=\delta_\xb$, we get
\be\label{dirder}
F_{C,\mu}(\xi;\delta_\xb) = 2 \left[ P_{C,\xi}(\xb)-P_{C,\mu}(\xb) - \SE_C(\xi)+\SE_C(\xi,\mu) \right] \,,
\ee
where $\SE_C(\xi,\nu)=\int_{\SX^2} C(\xb,\xb')\,\xi(\dd\xb)\nu(\dd\xb')$ and
\bea
P_{C,\xi}(\xb) = \int_\SX C(\xb,\xb')\,\xi(\dd\xb')
\eea
(respectively, $P_{C,\mu}(\xb) = \int_\SX C(\xb,\xb')\,\mu(\dd\xb')$) is called the potential of $\xi$ (respectively, of $\mu$), at $\xb$, associated with $C$. $P_{C,\mu}(\cdot)$ is also called the kernel embedding of $\mu$ is the RKHS associated with $C$ \citep[Def.~9]{SejdinovicSGF2013}. 
Standard kernel-herding corresponds to the Frank-Wolfe conditional gradient algorithm \citep{BachLJO2012}, that is, to the vertex-direction method with predefined step-length, commonly used in optimal experimental design since the pioneering work of \citet{Wynn70} and \citet{Fedorov72}.

The general form of the algorithm, with step length $\ma_k\in(0,1)$ at iteration $k$, is as follows: starting with some probability measure $\zeta^{(k_1)}$ on $\SX$, we take, for all $k\geq k_1$,
\be\label{kh}
\zeta^{(k+1)} = (1-\ma_k)\, \zeta^{(k)} + \ma_k\, \delta_{\zb_{k+1}}
\ee
where $\zb_{k+1} \in\Arg\min_{\zb\in\SX} F_{C,\mu}(\zeta^{(k)};\delta_\xb)$. From \eqref{dirder}, this is equivalent to
\be\label{z-n+1}
\zb_{k+1} \in\Arg\min_{\zb\in\SX} P_{C,\zeta^{(k)}}(\zb)-P_{C,\mu}(\zb) \,.
\ee
If the initial measure $\zeta^{(k_1)}$ is finitely supported on a set $\SS^{(k_1)}$, then $\zeta^{(k)}$ remains finitely supported for all $k$. Selecting the optimal  $\ma_k$ at each iteration corresponds to Fedorov's\nocite{Fedorov72} algorithm (1972) used in optimal design for parametric models. If $\SS^{(k_1)}$ has $k_1$ elements, $\SS^{(k_1)}=\Zb^{(k_1)}=\{\zb_1,\ldots,\zb_{k_1}\}$, and $\zeta^{(k_1)}= (1/k_1) \sum_{i=1}^{k_1} \delta_{\zb_i}$ is uniform on $\Zb^{(k_1)}$, by choosing $\ma_k=1/(k+1)$ for all $k\geq k_1$ we obtain that
$\zeta^{(k)}= (1/k) \sum_{i=1}^k \delta_{\zb_i}$ for all $k$, and
\bea
P_{C,\zeta^{(k)}}(\xb) = \frac{1}{k} \sum_{i=1}^k C(\xb,\zb_i) \,.
\eea
In particular, we can take $\ma_k=1/(k+1)$, $\zeta^{(1)}=\delta_{\zb_1}$ for some $\zb_1\in\SX$, and $\zb_1$ can be chosen by maximizing $P_{C,\mu}(\zb)$. For stationary kernels such that $C(\xb,\xb')$ only depends on $\xb'-\xb$, it amounts at taking $\zb_1$ at the center of $\SX$.

We shall denote by $\Zb_k=\KH(\Zb_{k_1},C,k)$ the $k$-point design design obtained in this way, after $k$ iterations of kernel herding initialized at $\Zb_{k_1}$ containing $k_1$ elements, with $\ma_k=1/(k+k_1)$ for all $k\geq 1$; $\KH(\emptyset,C,k)$ selects $\zb_1$ by maximization of $P_{C,\mu}$.

\vsp
In practice, the search for $\zb_{k+1}$ in \eqref{z-n+1} is generally made within a finite subset $\SX_Q$ of $\SX$, with $Q$ elements.
The cost of the determination of $\zb_{k+1}$ in \eqref{z-n+1} is $\SO(Q)$ if we compute $C(\xb,\zb_k)$ for all $\xb\in\SX_Q$ and update the sum $\sum_{i=1}^{k-1} C(\xb,\zb_i)$; the cost for $k$ iterations then scales as $\SO(k Q)$, including the initial cost for the computation of $P_{C,\mu}(\xb)$ for all $\xb\in\SX_Q$. Another option when $\mu$ is approximated by the uniform measure $\mu_Q$ on $\SX_Q$ and $C(\xb,\xb')$ only depends on $\|\xb-\xb'\|$, is to compute in advance the $Q(Q-1)/2$ distances between all pairs of points in $\SX_Q$ (feasible only if $Q$ is not too large).

\vsp
The minimum-norm variant of \citet{BachLJO2012} replaces $\zeta^{(k)}$ in \eqref{z-n+1} by the measure having the same support $\SS^{(k)}$ but optimal weights, positive and summing to one; these optimal weights are solution of a convex quadratic programming problem. Here we shall  consider a simplified version where $\zeta^{(k)}$ is replaced by $\hat\zeta^{(k)}$ having weights $\hat w_i^{(k)}$ summing to one and such that such that $\SE_C(\hat\zeta^{(k)}-\mu)$ is minimal. For a measure $\zeta_k$ with support $\SS^{(k)}$ and weights $\wb^{(k)}$, we have
\be\label{MMD2-w}
\SE_C(\zeta_k-\mu) =  {\wb^{(k)}}\TT \Cb_k \wb^{(k)} - 2\, {\wb^{(k)}}\TT \pb_{C,k}(\mu) + \SE_C(\mu) \,,
\ee
where $\{\Cb_k\}_{i,j}=C(\zb_i,\zb_j)$, $i,j=1\ldots,k$ and $\pb_{C,k}(\mu)=\left[P_{C,\mu}(\zb_1),\ldots,P_{C,\mu}(\zb_k)\right]\TT$.
Its minimization under the constraint $\1b_k\TT\wb^{(k)}=1$, with $\1b_k$ the $k$-dimensional vector with all components equal to one, gives the optimal weights
\be\label{wopt-sum=1}
\hat \wb^{(k)} = (\hat w_1^{(k)},\ldots,\hat w_k^{(k)})\TT = \left(\Cb_k^{-1} - \frac{\Cb_k^{-1} \1b_k\1b_k\TT \Cb_k^{-1}}{\1b_k\TT\Cb_k^{-1}\1b_k} \right) \pb_{C,k}(\mu) + \frac{\Cb_k^{-1} \1b_k}{\1b_k\TT\Cb_k^{-1}\1b_k} \,.
\ee
By construction, $\hat\zeta^{(k)}$ minimizes $\SE_C(\zeta_k-\mu)$ with respect to measures $\zeta_k$ of total mass one supported on $\SS^{(k)}$, and one can show \citep{PZ2020-SIAM} that its potential $P_{C,\hat\zeta^{(k)}}(x)$ satisfies
\bea
P_{C,\hat\zeta^{(k)}}(x) - P_{C,\mu}(x) - \SE_C(\hat\zeta^{(k)},\mu)+\SE_C(\mu) = 0\,, \ \forall x\in\SS^{(k)} \,,
\eea
showing that $P_{C,\hat\zeta^{(k)}}(x) - P_{C,\mu}(x)$ is constant on $\SS^{(k)}$.

When $\SX$ is discretized into $\SX_Q$, the substitution of $\hat\zeta^{(k)}$ for $\zeta^{(k)}$ requires the storage of all $C(\xb,\zb_i)$, $i=1,\ldots,k$, $\xb\in\SX_Q$, in order to compute $P_{C,\hat\zeta^{(k)}}(\xb)=\sum_{i=1}^k \hat w_i^{(k)} C(\xb,\xb_i)$ in \eqref{z-n+1}. At iteration $k$, the computation of $\hat \wb^{(k)}$ by \eqref{wopt-sum=1} also induces an additional computational cost of $\SO(k^3)$  (reduced to $\SO(k^2)$ if rank-one updating is used to compute $\Cb_k^{-1}$); $\mg_C^2(\hat\zeta^{(k)},\mu)$ decreases faster than $\mg_C^2(\zeta^{(k)},\mu)$; see \citet{P2021}.

We shall denote by $\Zb_k=\MN(\Zb_{k_1},C,k)$ the $k$-point design obtained after $k$ iterations, initialized at $\Zb_{k_1}$  ($\MN(\emptyset,C,k)$ chooses $\zb_1$ that maximizes $P_{C,\mu}(\zb)$). We write $[\Zb_k,\hat\wb^{(k)}]=\MN(\Zb_{k_1},C,k)$ when we are also interested in the weights $\hat\wb^{(k)}$ given by \eqref{wopt-sum=1}, and for any $m$-point design $\Zb_m$ we denote by $\hat\wb(\Zb_m,C)$ the weights computed by \eqref{wopt-sum=1}.

\paragraph{Example 1.}
To illustrate the behavior of the algorithms above, we consider a small one-dimensional example with $\SX=[0,1]$ and $C=K_{3/2,\mt}$, the Mat\'ern $3/2$ kernel
\be\label{K32}
K_{3/2,\mt}(x,x')=(1+\sqrt{3}\,\mt\,|x-x'|)\, \exp(-\sqrt{3}\,\mt\,|x-x'|) \,. 
\ee
The measure $\mu$ is approximated by the uniform discrete distribution on $\SX_Q$ given by the first $Q=256$ points of a scrambled Sobol' sequence in $\SX$; $\zb_{k+1}$ in \eqref{z-n+1} is searched within the same set $\SX_Q$; we take $\mt=10$ in $K_{3/2,\mt}$.

\begin{figure}[ht!]
\begin{center}
\includegraphics[width=.49\linewidth]{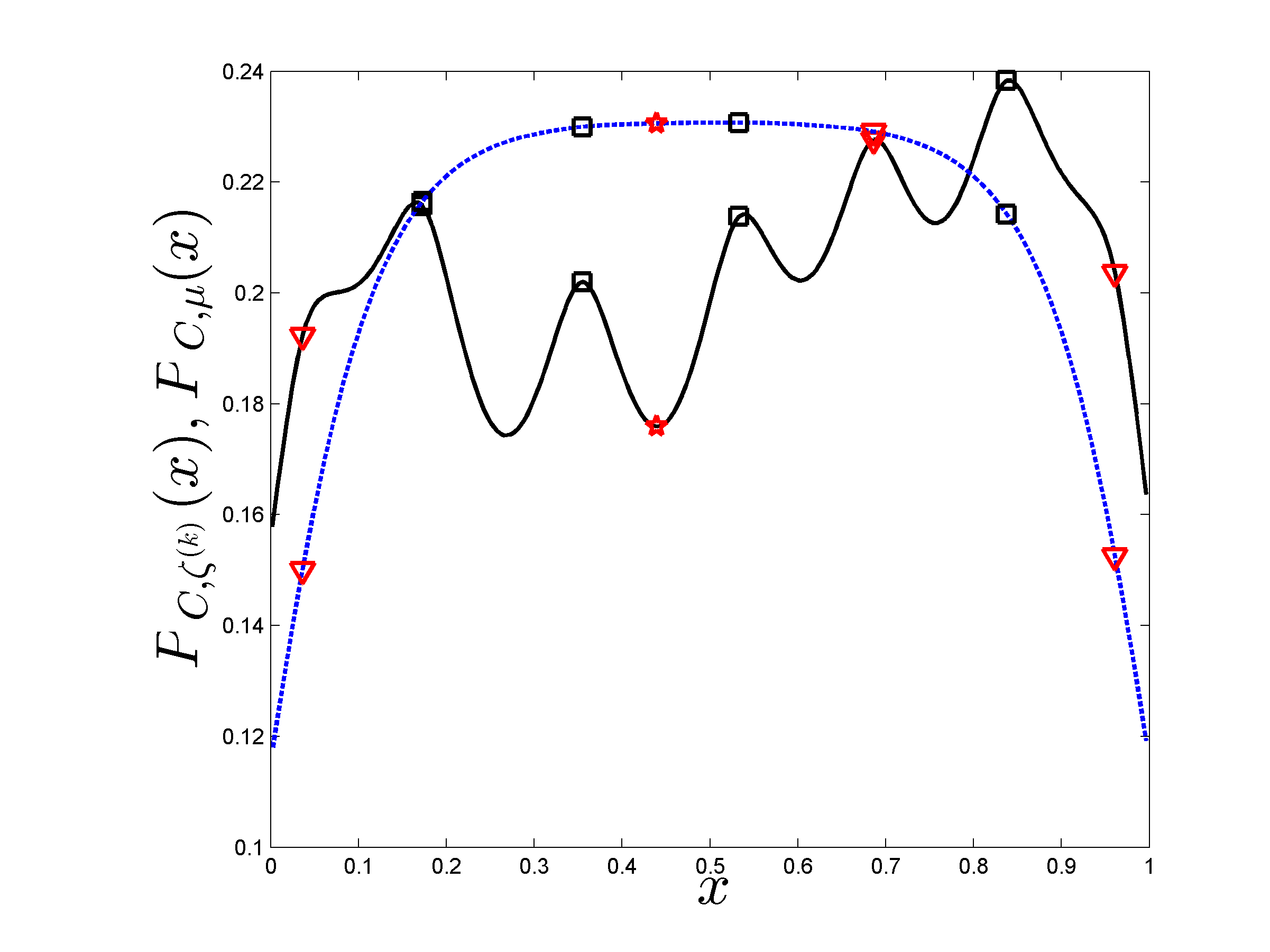} \includegraphics[width=.49\linewidth]{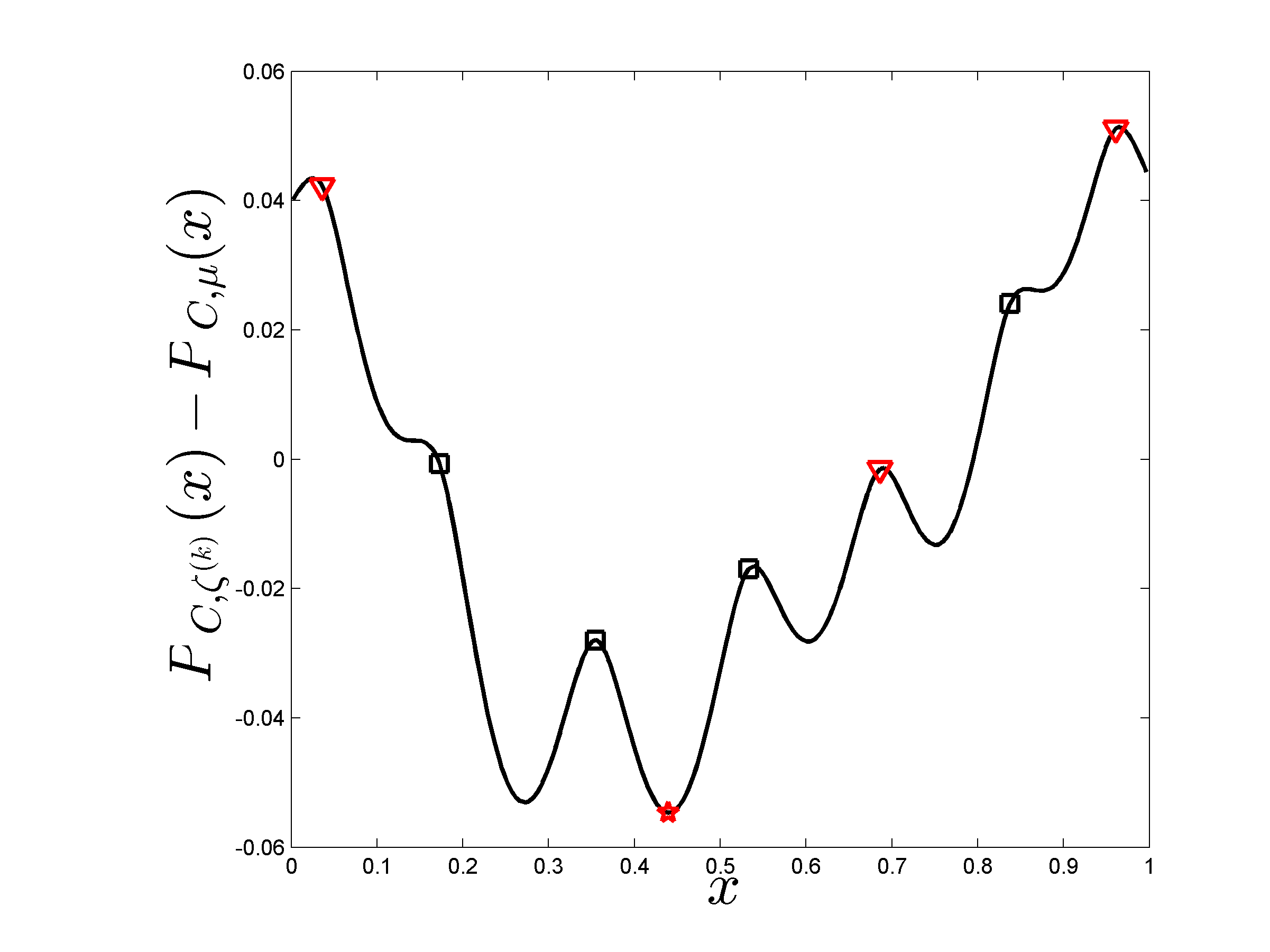}

\includegraphics[width=.49\linewidth]{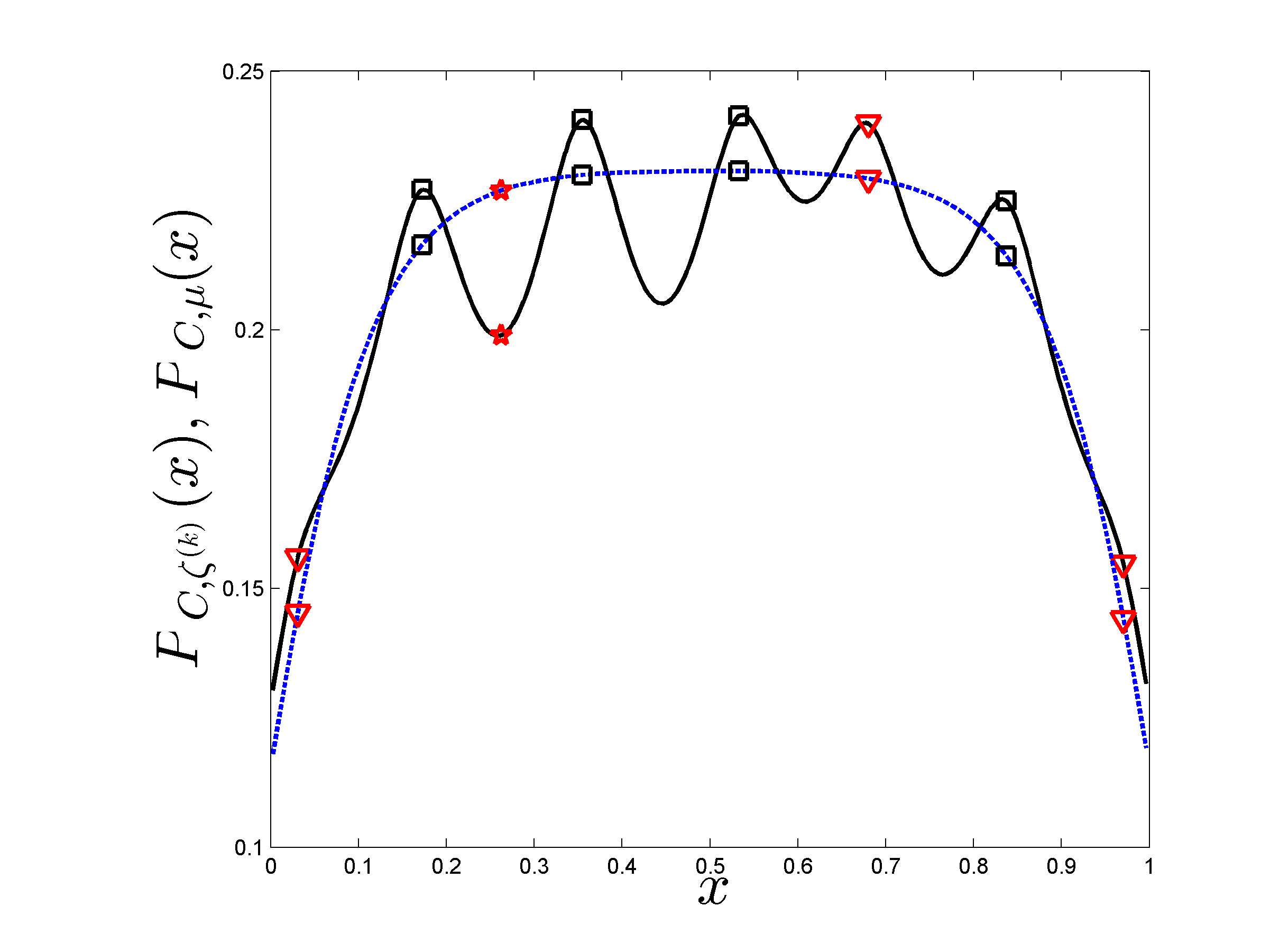} \includegraphics[width=.49\linewidth]{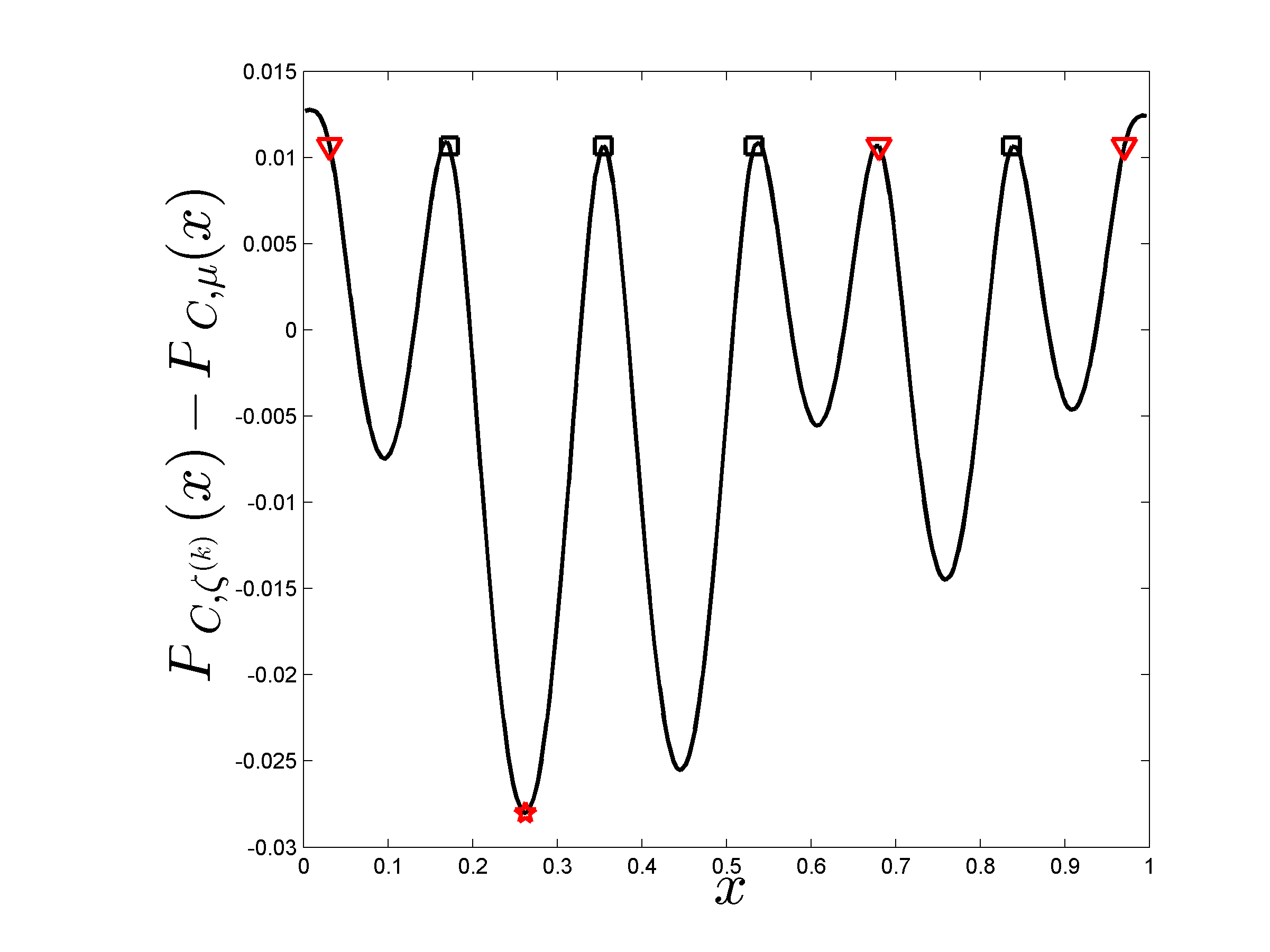}

\end{center}
\caption{\small Left: $P_{C,\zeta^{(k)}}(x)$ (black solid line) and $P_{C,\mu}(x)$ (blue dashed line); Right: $P_{C,\zeta^{(k)}}(x)-P_{C,\mu}(x)$. Design points $x_i$, $i=1,\ldots,n=4$: black $\square$; validation points $z_i$, $i\leq k=3$: red $\triangledown$; $z_{4}$: red $\bigstar$. Top row: kernel herding; bottom row: MN variant; $C=K_{3/2,\mt}$.}
\label{F:Pxi&Pmu_KH_d1_n4_m4_M32}
\end{figure}

\vsp
The left panel of Figure~\ref{F:Pxi&Pmu_KH_d1_n4_m4_M32} shows $P_{C,\zeta^{(k)}}(x)$ (black solid line) and $P_{C,\mu}(x)$ (blue dashed line) as functions of $x\in\SX$, with $\Xb_n=\KH(\emptyset,C,n)$; the right panel shows $P_{C,\zeta^{(k)}}(x)-P_{C,\mu}(x)$. The figure is for $n=4$ and $\Xb_n$ is indicated by black squares.

Kernel herding is used on the top row: $\zeta^{(k)}$ for $k=3$ is supported by the points in $\KH(\Xb_n,C,3)$ indicated with a red triangle; the next point $z_{4}$ chosen by the algorithm --- the
location of minimum of the right panel --- is indicated by the red star. $P_{C,\zeta^{(k)}}(x)$ decreases when $x$ moves away from its closest prediction point $x_i$ or validation point $z_i$, whereas $P_{C,\mu}(\cdot)$ is a fixed function, independent of the $x_i$ and $z_i$. The bottom row is for the minimum-norm variant $\MN(\Xb_n,C,k)$ of kernel herding: at iteration $k$ we replace $\zeta^{(k)}$ by $\hat\zeta^{(k)}$ having weights given by \eqref{wopt-sum=1}. The right panel illustrates the property that $P_{C,\hat\zeta^{(k)}}(x) - P_{C,\mu}(x)$ is constant on the support $\SS^{(k)}=\Xb_n \cup \{z_1,\ldots,z_k\}$ of $\hat\zeta^{(k)}$. Note that $P_{C,\hat\zeta^{(k)}}(x)$ is closer to $P_{C,\mu}(x)$ than in the first row, indicating a better approximation of $\mu$ in the sense of the MMD criterion.
\fin

To summarise, we proposed two distinct validation designs in this section:
\begin{itemize}
   \item The kernel herding solution $\Zb_k = \KH(\Zb_{k_1},C, k)$, a design of size $k$ obtained by iteratively minimising $P_{C,\zeta^{(k)}}(\zb) - P_{C,\mu}(\zb)$, see \eqref{z-n+1}, starting from $\Zb_{k_1}$ of size $k_1$, and with weights $\ma_k = 1/(k + k_1)$ for all $k \geq 1$. A variant of this notation will be introduced in the next subsection, where we will introduce $\Zb_m = \KH(\Zb_{k_1}, C, k, \setminus \Xb_n)$ to denote the design of size $m$, with no repeated points and empty intersection with $\Xb_n$, obtained after $k$ iterations.
   \item The minimum norm solution $\Zb_k = \MN(\Zb_{k_1}, C, k)$, a design of size $k$ obtained after $k$ iterations of \eqref{z-n+1} initiated at $\Zb_{k_1}$, but using a measure $\zeta^{(k)}$ with the optimal weights \eqref{wopt-sum=1} in the computation of $P_{C,\zeta^{(k)}}(\zb)$. As we will see below, this solution is not well defined when $C=K_{|n}$ or $C=\bK_{|n}$, and a slightly different definition, dropping the constraint of unitary sum of the weights of $\zeta^{(k)}$, is required. It will be denoted by $\Zb_k = \MN_2(\Zb_{k_1}, C, k)$.
 \end{itemize}

\subsection{Incremental construction of space-filling and validation designs}\label{S:kh-validation}

To apply a kernel-herding algorithm to the minimization of $\mg_{\bK_{|n}}(\zeta_m,\mu)$ given by \eqref{delta-MMD} with respect to $\zeta_m$, we simply substitute the conditional kernel $\bK_{|n}$, given by \eqref{bK_|n}, for the kernel $C$ in (\ref{kh}, \ref{z-n+1}).
The construction has the advantage of being incremental\footnote{However, it does not provide the optimal design for $m$ fixed: the construction of one-shot $m$-point designs minimizing a MMD criterion is considered for instance in \citep{PZ2020-SIAM}; we do not develop this aspect here.}: it generates a design sequence $\zb_1,\zb_2,\ldots$ which can be interrupted at any design size $m$.

However, the application of kernel herding to kernels $C$ such that $C(\xb,\xb_i)=0$ for all design points $\xb_i$, which occurs when  $C=\bK_{|n}$, requires a specific treatment.  
Indeed, it may then happen that the two potentials $P_{C,\zeta^{(k)}}(\xb_i)$ and $P_{C,\mu}(\xb_i)$ used in \eqref{z-n+1} satisfy
$P_{C,\mu}(\xb)\leq P_{C,\zeta^{(k)}}(\xb)$ for all $\xb\in\SX$, with $P_{C,\mu}(\xb)= P_{C,\zeta^{(k)}}(\xb)$ for $\xb\in\Xb_n$ and the inequality being strict otherwise. In that case, \eqref{z-n+1} necessarily chooses $\zb_{k+1}$ among $\Xb_n$. If a $\zeta^{(\ell)}$ has one of its support points $\zb_j$ in $\Xb_n$, \eqref{MMD2-w} indicates that the associated weight $w^{(\ell)}_j$ does not contribute to $\SE_C(\zeta^{(\ell)}-\mu)$. The selection of $\zb_{\ell+1}$ among $\Xb_n$ is thus equivalent to a reduction of the total mass of other points that contribute to $\SE_C(\zeta^{(\ell)}-\mu)$.

The possible selection of $\zb_{k+1}$ within $\Xb_n$ has several consequences on kernel herding.
\begin{description}
  \item[(\textit{i})] When it happens that $\zb_{k+1}$ is chosen among $\Xb_n$ at an iteration \eqref{z-n+1} of standard kernel herding (with uniform weighting), we can nevertheless continue iterations until the number of selected points not in $\Xb_n$ reaches the desired value $m$; we denote by $\Zb_m=\KH(\Zb_{k_1},C,k,m,\setminus\Xb_n)$ the corresponding $m$-point design.
      Since the selection is made within a finite set, it may also happen that the same point is selected several times. In that case, we may also impose that $\Zb_m$ contains $m$ distinct points and continue iterations until this condition is satisfied; the weights given by the algorithm to the $\zb_i$ in $\Zb_m$ are then multiple of $1/m'$, with $m'\geq m$ (they are not necessarily all equal to $1/m$).
  \item[(\textit{ii})] When a support point $\zb_j$ of $\zeta^{(k)}$ coincides with a design points $\xb_i$, $\Cb_k$ is singular and we cannot compute $\hat \wb^{(k)}$ by \eqref{wopt-sum=1}; that is, the minimum-norm variant of kernel herding cannot be used.
  \item[(\textit{iii})]  When a support point $\zb_j$ belongs to $\Xb_n$, the optimal weights allocated to the points that do not belong to $\Xb_n$ are obtained by minimizing \eqref{MMD2-w} with respect to $\wb^{(k)}$ \emph{without the constraint $\1b_k\TT\wb^{(k)}=1$}.
\end{description}

To account for the possibility that the algorithm may choose $\zb_{k+1}$ in $\Xb_n$, we consider a new version of kernel herding where, at iteration $k$, $\zeta^{(k)}$ is replaced by $\breve\zeta^{(k)}$ having the same support $\SS^{(k)}$ but weights $\breve\wb^{(k)}$ that minimize $\SE_C(\zeta^{(k)}-\mu)$ given by \eqref{MMD2-w} with respect to $\wb^{(k)}$ without constraints on $\wb^{(k)}$, contrarily to $\hat\wb^{(k)}$, given by \eqref{wopt-sum=1}, which satisfies $\sum_{i=1}^k \hat w_i^{(k)}=1$. Direct calculation gives
\be\label{newww}
\breve\wb^{(k)} = (\breve w^{(k)}_1,\ldots,\breve w^{(k)}_k)\TT = \Cb_k^{-1}\pb_{C,k}(\mu) \,.
\ee
The measures $\zeta^{(k)}$, $\hat\zeta^{(k)}$ and $\breve\zeta^{(k)}$, with respective weights $\1b_k$, $\hat\wb^{(k)}$ and $\breve\wb^{(k)}$, satisfy
\bea
\SE_C(\zeta^{(k)}-\mu) \geq \SE_C(\hat\zeta^{(k)}-\mu) \geq \SE_C(\breve\zeta^{(k)}-\mu) \,.
\eea
In general, both $\hat\wb^{(k)}$ and $\breve\wb^{(k)}$ may have negative components.
We shall denote by $\Zb_k=\MN_2(\Zb_{k_1},K,k)$ the design obtained after $k$ iterations of this variant of kernel herding, initialized at $\Zb_{k_1}$ ($\MN_2(\emptyset,K,k)$ chooses $\zb_1$ that maximizes $P_{K,\mu}$).
We write $[\Zb_k,\breve\wb^{(k)}]=\MN_2(\Zb_{k_1},K,k)$ when we are also interested in the weights $\breve\wb^{(k)}$ given by \eqref{newww}, and for any $m$-point design $\Zb_m$, we denote by $\breve\wb(\Zb_m,K)$ the weights computed by \eqref{newww}.

This construction can be interpreted as standard kernel herding applied to a kernel $C(k)$ varying along iterations, given by the conditional kernel $C_{|k}$ at iteration $k$,
\bea
C_{|k}(\xb,\xb')=C(\xb,\xb')-\cb_k\TT(\xb)\Cb_k^{-1}\cb_k(\xb') \,,
\eea
where $\cb_k(\xb) = \left[C(\xb,\zb_1)\,\ldots,C(\xb,\zb_k)\right]\TT$, see \eqref{covn}. Indeed, the potential $P_{C_{|k},\zeta}(\zb)$ for a measure $\zeta$ on $\SX$ is
\be\label{P-cond}
P_{C_{|k},\zeta}(\zb) = P_{C,\zeta}(\zb) - \cb_k\TT(\zb)\Cb_k^{-1}\pb_{C,k}(\zeta)\,,
\ee
where $\pb_{C,k}(\zeta)=\left[P_{C,\zeta}(\zb_1),\ldots,P_{C,\zeta}(\zb_k)\right]\TT$. Since $C_{|k}(\zb,\zb_i)=0$ for all $\zb_i$, for any $\zeta_k$ supported on $\Zb_k=\{\zb_1,\ldots,\zb_k\}$ we have, for all $\zb\in\SX$,
\be
P_{C_{|k},\zeta_k}(\xb) - P_{C_{|k},\mu}(\zb) = - P_{C_{|k},\mu}(\zb) &=& \cb_k\TT(\zb)\Cb_k^{-1}\pb_{C,k}(\mu) - P_{C,\mu}(\zb) \label{delta-P} \\
&=& P_{C,\breve\zeta^{(k)}}(\zb)-P_{C,\mu}(\zb) \,. \label{sameP}
\ee
At iteration $k$, kernel herding with $C_{|k}$ and the variant with kernel $C$ but optimal weights $\breve\wb^{(k)}$ thus select the same
$\zb_{k+1}$ that minimizes \eqref{sameP}. Note that $P_{C,\breve\zeta^{(k)}}(\zb)-P_{C,\mu}(\zb)=0$ for all $\zb_i$.
When we substitute the conditional kernel $K_{|n}$ for $C$, the variant $\MN_2$ of kernel herding also satisfies the following property, meaning that we do not need to know where the points $\Xb_n$ are, everything in terms of information being coded in the conditional kernel $K_{|n}$.

\begin{thm}\label{Th:1} For any positive definite kernel $K$, any design $\Xb_n$ and any $k\geq 1$, there exist choices for $\zb_{i+1}$, $i=0,\ldots,k$ in \eqref{z-n+1} such that $\MN_2(\Xb_n,K,k)=\MN_2(\emptyset,K_{|n},k)$, where $K_{|n}$ is defined by \eqref{covn}.
\end{thm}

\noindent{\em Proof.} Consider first the case $k=1$. On the one hand, $\zb_1=\MN_2(\Xb_n,K,1)$ minimizes $\kb_n\TT(\zb)\Kb_n^{-1}\pb_{K,n}(\mu)-P_{K,\mu}(\zb)$, see \eqref{delta-P}; on the other hand, $\zb'_1=\MN_2(\emptyset,K_{|n},1)$ maximizes $P_{K_{|n},\mu}(\zb)=P_{K,\mu}(\zb) - \kb_n\TT(\zb)\Kb_n^{-1}\pb_{K,n}(\mu)$, see \eqref{P-cond}. One can therefore choose $\zb_1=\zb'_1$.

The identity of the two constructions at any $k>1$ is a consequence of the conditioning property of GP: at step $k$, they both use the kernel $K_{|n+k}$. More precisely, $\zb_{k+1}$ for the construction of $\MN_2(\Xb_n,K,k+1)$ minimizes
$J(\zb) = \kb_{n+k}\TT(\zb)\Kb_{n+k}^{-1}\pb_{K,n+k}(\mu)-P_{K,\mu}(\zb)$, where
\bea
\kb_{n+k}(\zb) = \left(
                   \begin{array}{c}
                     \kb_{n}(\zb) \\
                     \kb_{k}(\zb) \\
                   \end{array}
                 \right)\,, \
                 \pb_{K,n+k}(\mu) = \left(
                   \begin{array}{c}
                     \pb_{K,n}(\mu)\\
                     \pb_{K,k}(\mu) \\
                   \end{array}
                 \right)\,, \
                 \Kb_{n+k} = \left(
                               \begin{array}{cc}
                                 \Kb_{n} & \Kb_{n,k} \\
                                 \Kb_{k,n} & \Kb_{k} \\
                               \end{array}
                             \right)\,,
\eea
while $\zb'_{k+1}$ for $\MN_2(\emptyset,K_{|n},k+1)$ minimizes $J'(\zb) =
{\kb_{|n}}_k\TT(\zb){\Kb_{|n}}_k^{-1}\pb_{K_{|n},k}(\mu)-P_{K_{|n},\mu}(\zb)$, where
\bea
{\kb_{|n}}_k(\zb)=\kb_k(\zb)-\Kb_{k,n}\Kb_n^{-1}\kb_n(\zb)\,, \ \pb_{K_{|n},k}(\mu)=\pb_{K,k}(\mu)-\Kb_{k,n}\Kb_n^{-1}\pb_{K,n}(\mu)
\eea
and ${\Kb_{|n}}_k = \Kb_k-\Kb_{k,n}\Kb_n^{-1}\Kb_{n,k}$. Direct application of Woodbury identity for matrix inversion and inversion of a block matrix shows that $J(\zb)=J'(\zb)$; we can thus choose the same $\zb_{k+1}$ in both constructions in case multiple choices are possible.
\carre

\paragraph{Example 1 (continued).}

We consider the same situation as in Example~1 with the same $\Xb_n=\KH(\emptyset,K,n)$ for $K=K_{3/2,10}$.

\begin{figure}[ht!]
\begin{center}

\includegraphics[width=.49\linewidth]{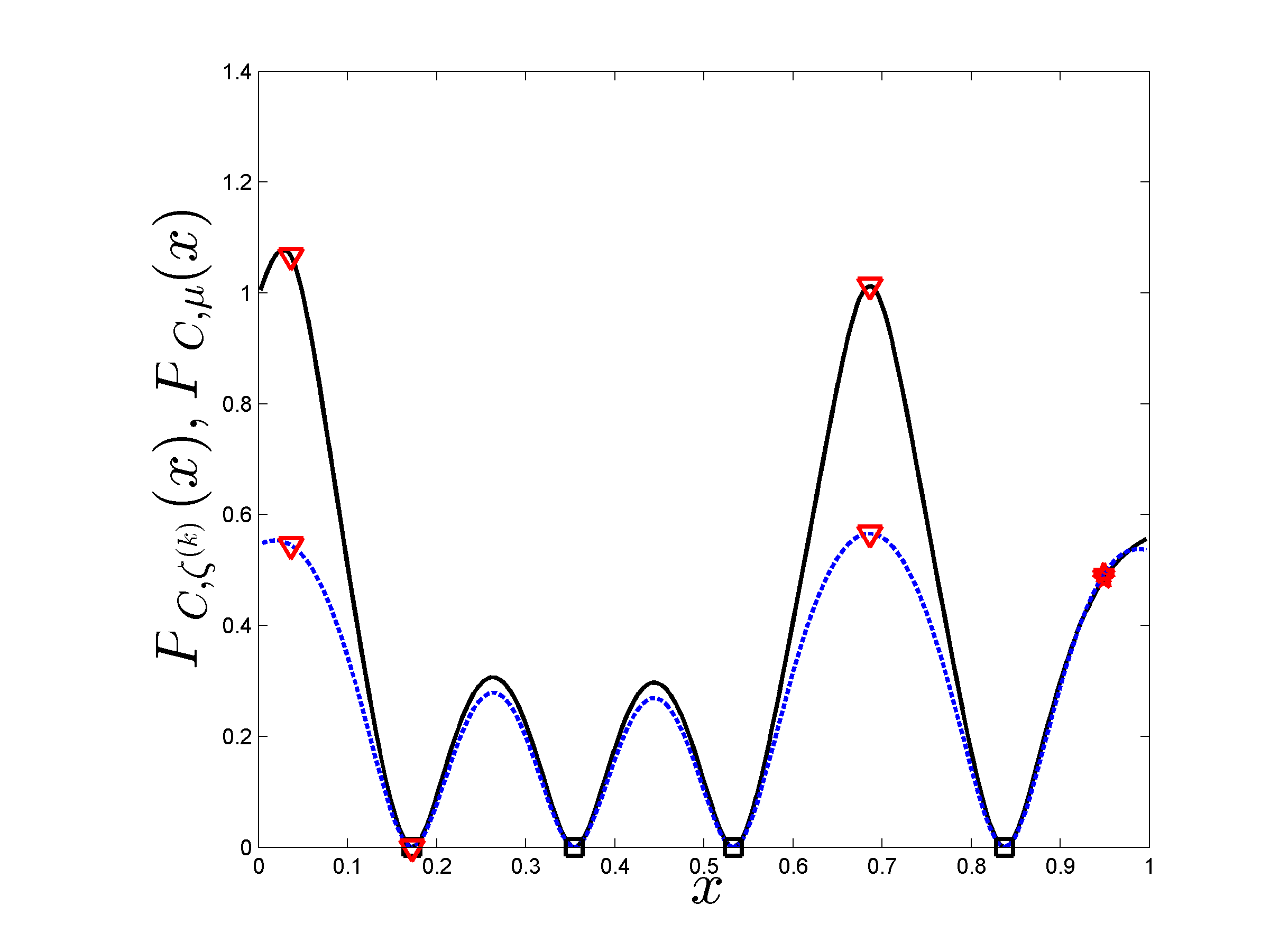} \includegraphics[width=.49\linewidth]{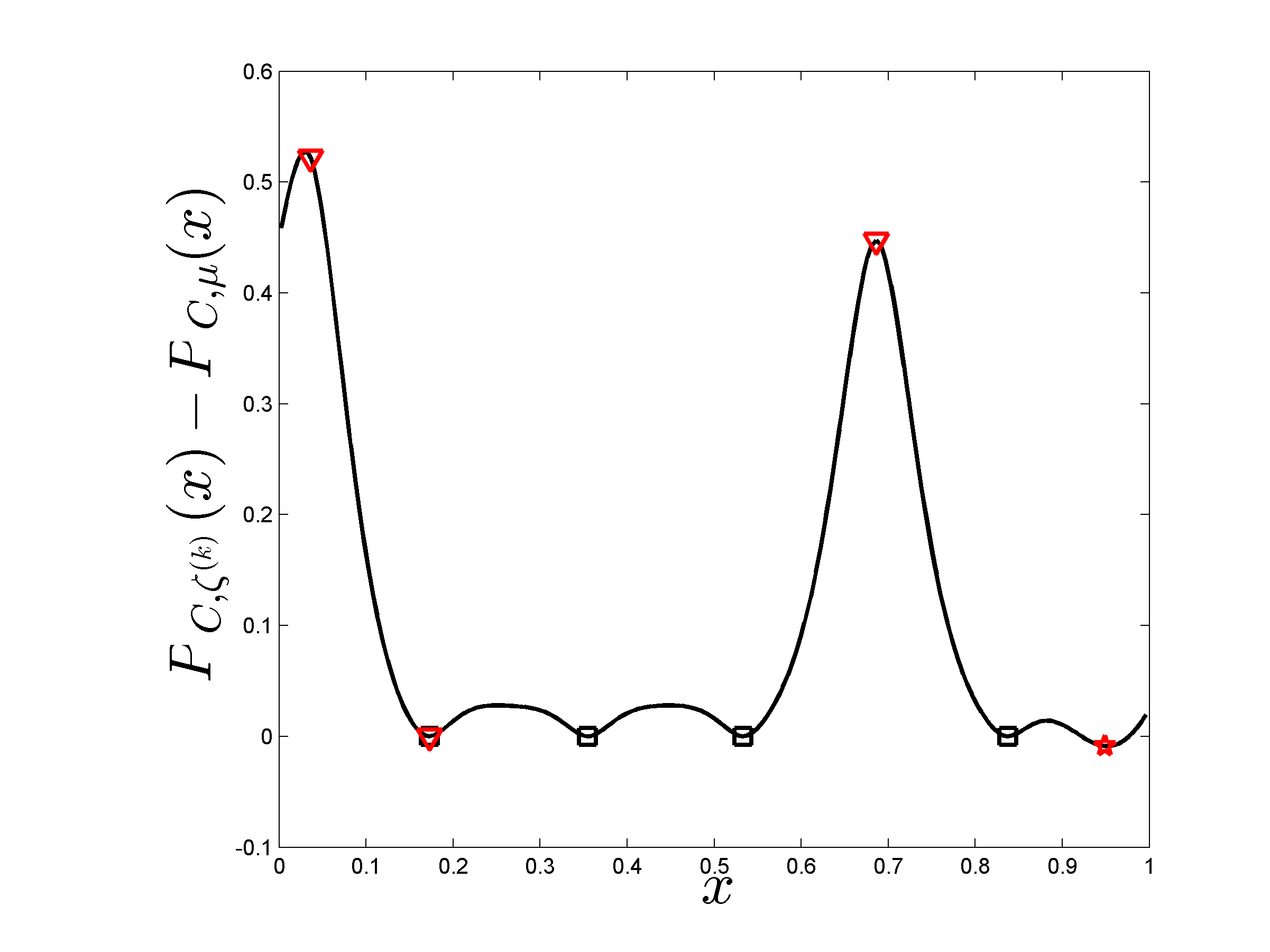}
\end{center}
\caption{\small Kernel herding $\KH(\emptyset,C,4)$ for $C=\bK_{|n}$. Left: $P_{C,\zeta^{(k)}}(x)$ (black solid line) and $P_{C,\mu}(x)$ (blue dashed line); Right: $P_{C,\zeta^{(k)}}(x)-P_{C,\mu}(x)$. Design points $x_i$, $i=1,\ldots,n=4$: black $\square$; validation points $z_i$, $i\leq k=3$: red $\triangledown$; $z_{4}$: red $\bigstar$.}
\label{F:Pxi&Pmu_KH_Cn2_d1_n4_m4_M32}
\end{figure}

Figure~\ref{F:Pxi&Pmu_KH_Cn2_d1_n4_m4_M32} illustrates the construction of $\KH(\emptyset,C,m)$ for $C=\bK_{|n}$, with $P_{C,\zeta^{(k)}}(x)$ (black solid line) and $P_{C,\mu}(x)$ (blue dashed line) as functions of $x\in\SX$ on the left and $P_{C,\zeta^{(k)}}(x)-P_{C,\mu}(x)$ on the right. $\Xb_n$ is indicated by black squares; $\zeta^{(k)}$ for $k=3$ is supported by the points indicated with a red triangle; the next point $z_{4}$ chosen by the algorithm corresponds to the red star. $P_{C,\zeta^{(k)}}(x_i)=P_{C,\mu}(x_i)$ for all $x_i\in\Xb_n$ and large values of potentials are obtained far away from the design points in $\Xb_n$ only. Note that $P_{C,\mu}(z)<P_{C,\zeta^{(3)}}(z)$ excepted in a small neighborhood around $z_4$, and that one of previous points selected by kernel herding (here $z_2$) coincides with a design point.

\begin{figure}[ht!]
\begin{center}
\includegraphics[width=.49\linewidth]{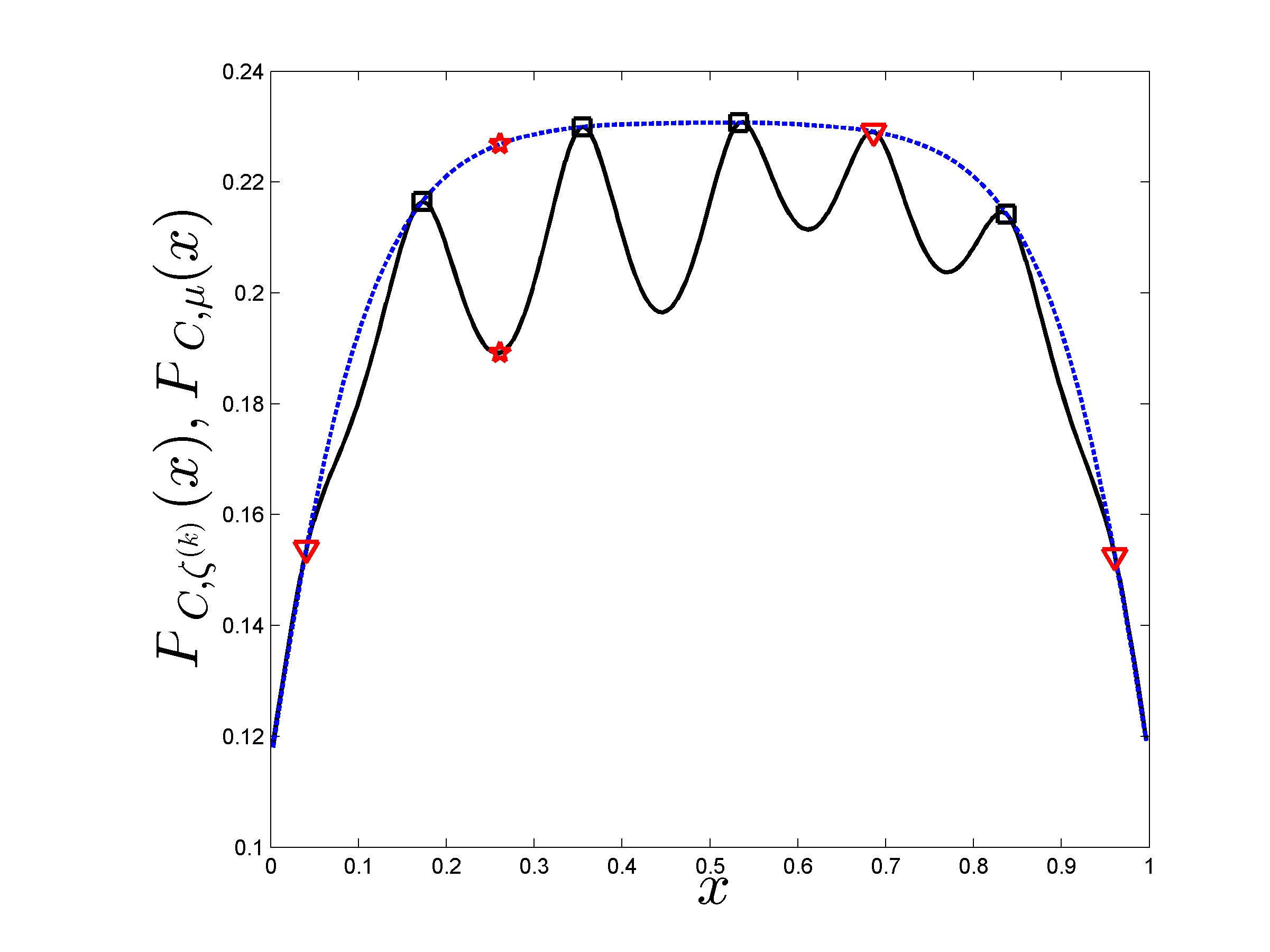} \includegraphics[width=.49\linewidth]{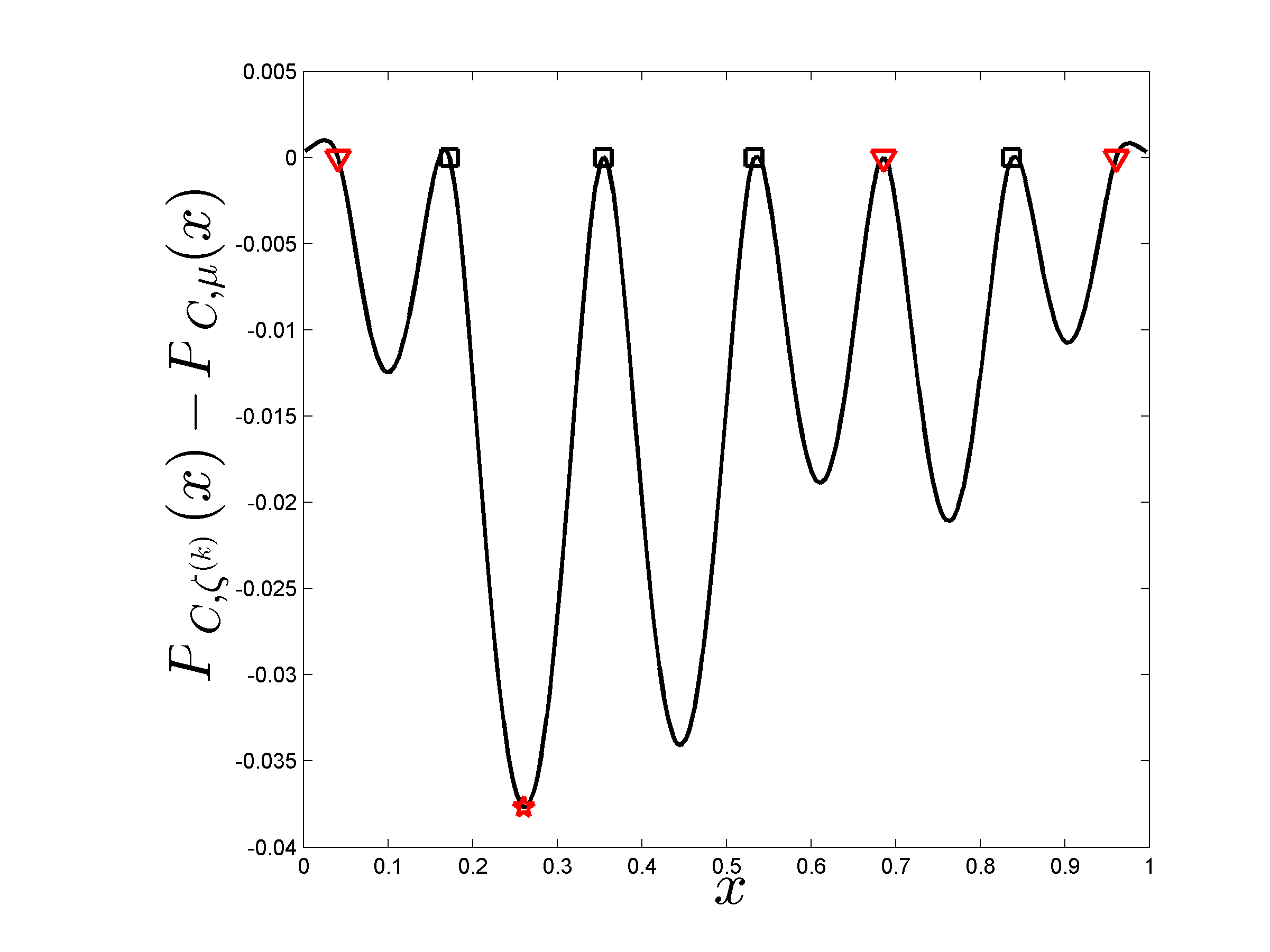}

\includegraphics[width=.49\linewidth]{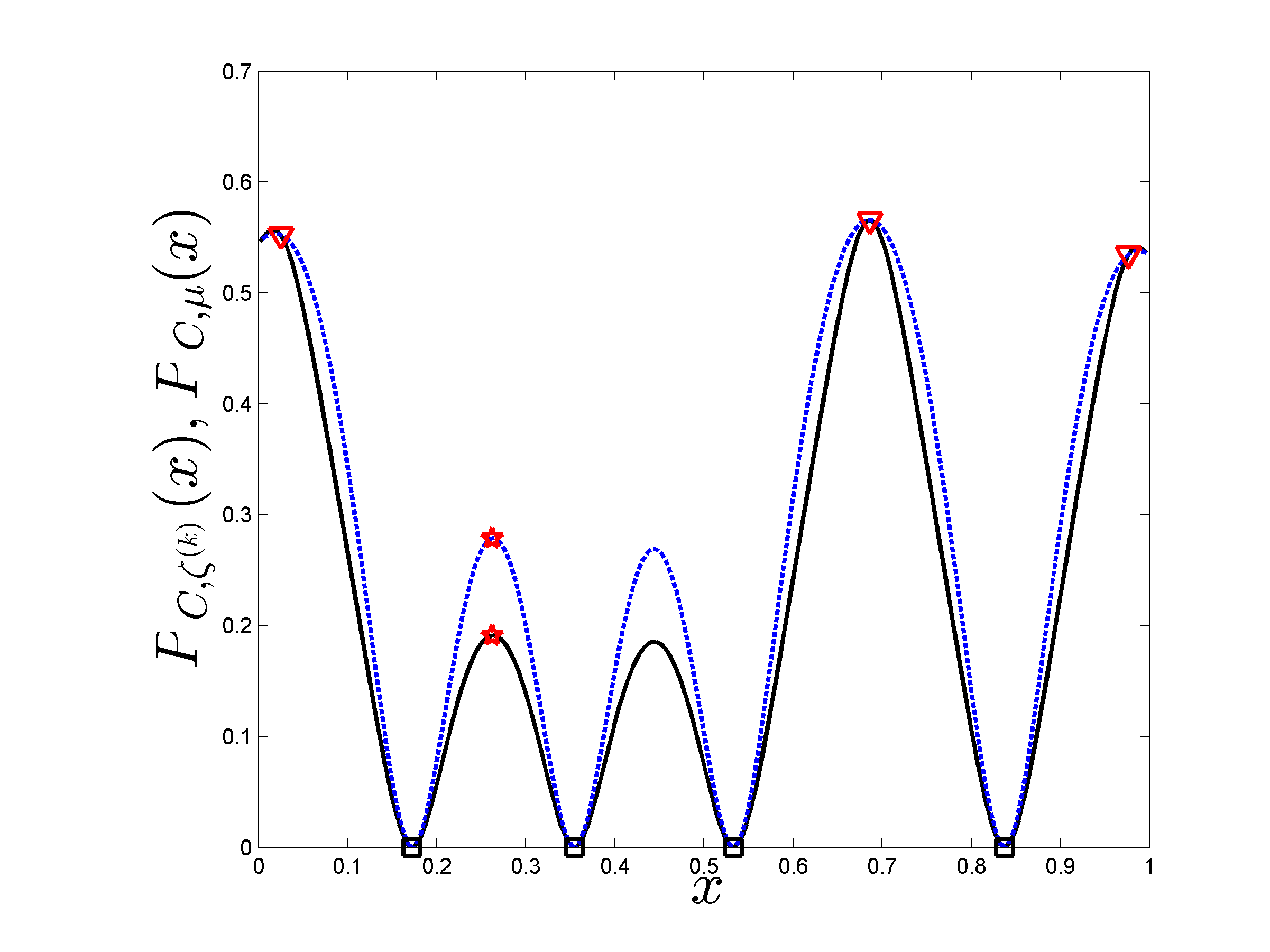} \includegraphics[width=.49\linewidth]{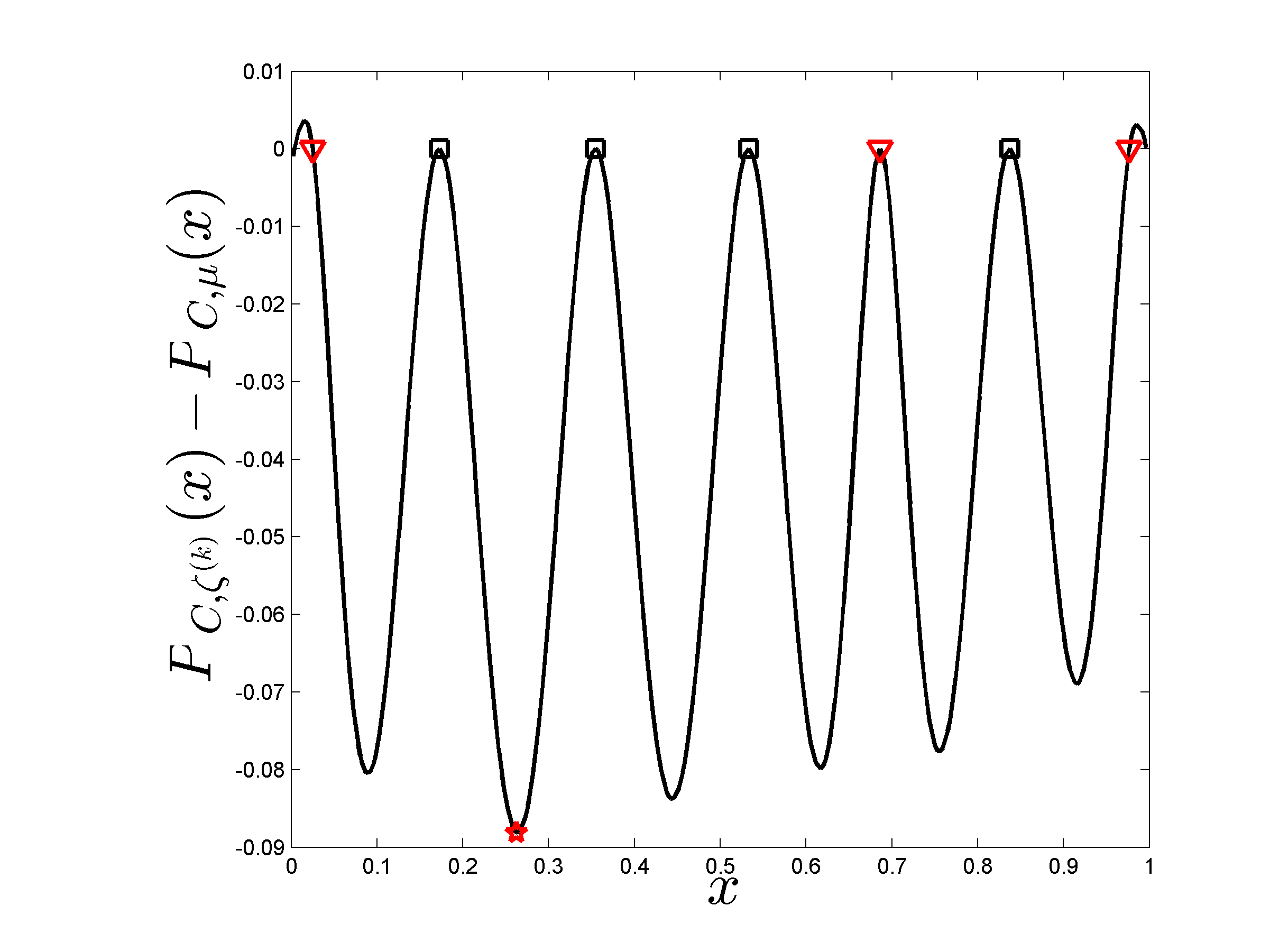}
\end{center}
\caption{\small $\MN_2$ variant of kernel herding. Left: $P_{C,\zeta^{(k)}}(x)$ (black solid line) and $P_{C,\mu}(x)$ (blue dashed line); Right: $P_{C,\zeta^{(k)}}(x)-P_{C,\mu}(x)$; $C=K$ (top row) and $C=\bK_{|n}$ (bottom row). Design points $x_i$, $i=1,\ldots,n=4$: black $\square$; validation points $z_i$, $i\leq k=3$: red $\triangledown$; $z_{4}$: red $\bigstar$.}
\label{F:Pxi&Pmu_MN2_d1_n4_m4_M32}
\end{figure}

We cannot use $\MN$ for $C=\bK_{|n}$ since we cannot compute optimal weights through \eqref{wopt-sum=1};
Figure~\ref{F:Pxi&Pmu_MN2_d1_n4_m4_M32} illustrates the construction with the minimum-norm variant $\MN_2$ of kernel herding that uses weights \eqref{newww}. In the first row, we use $C=K$ and at iteration $k$ the support $\SS^{(k)}$ equals $\Xb_n \cup \{z_1,\ldots,z_k\}$. The second row corresponds to $C=\bK_{|n}$.
Note that in both cases $P_{C,\breve\zeta^{(k)}}(\zb_i) = P_{C,\mu}(\zb_i)$ for all $\zb_i$ in the support $\SS^{(k)}$ of $\breve\zeta^{(k)}$.
\fin

\vsp
Although they follow the same principle of one-step ahead minimization of a convex functional of a measure, the three methods $\KH$, $\MN$ and $\MN_2$ rely on quite different functions $P_{C,\zeta^{(k)}}(x)-P_{C,\mu}(x)$ for the selection of support points in \eqref{z-n+1}; see the right columns of Figures~\ref{F:Pxi&Pmu_KH_d1_n4_m4_M32} to \ref{F:Pxi&Pmu_MN2_d1_n4_m4_M32}. The differences are also important depending on which kernel is used: the original one $K$, which is stationary in Example~1 above, or $\bK_{|n}$ which accounts for the presence of the $n$ design points in $\Xb_n$. Next section contains a numerical comparison of the performances of designs obtained with those different approaches, in particular in terms of
$\bDelta(\Zb_m,\Xb_n)$ given by
\eqref{delta-MMD}.

\section{Properties of validation design constructed by kernel herding}\label{S:Examples}

In this section, we investigate and compare the properties of validation designs obtained by minimizing $\mg_{\bK_{|n}}(\zeta_m,\mu)$ for different choices of $n$, $m$ and dimension $d$.
In the kernel-herding algorithm and its variants, we approximate $\mu$ by the uniform measure $\mu_Q$ on $\SX_Q$ given by the first $Q=2^{12}$ points of a scrambled Sobol' sequence in $\SX=[0,1]^d$; $Q$ is taken small enough the allow the computation of all $Q(Q-1)/2$ distances between pairs of points in $\SX_Q$. $K$ is the Mat\'ern $3/2$ isotropic kernel,
\bea 
K_{3/2,\mt}(\xb,\xb')=(1+\sqrt{3}\,\mt\,\|\xb-\xb'\|)\, \exp(-\sqrt{3}\,\mt\,\|\xb-\xb'\|) \,, 
\eea
with $\mt=n^{1/d}$ and $n$ the size of the prediction design $\Xb_n$, given by $\Xb_n=\KH(\emptyset,K,n)$. 

\paragraph{a) Space-filling performance.}
Although $\bDelta^2(\Zb_m,\Xb_n)= \mg_{\bK_{|n}}^2(\zeta_m,\mu)$ given by \eqref{delta-MMD} is not directly related to a space-filling characteristic, below we shall see that kernel herding applied to its minimization may provide designs with attractive space-filling properties. 

\begin{figure}[ht!]
\begin{center}
\includegraphics[width=.49\linewidth]{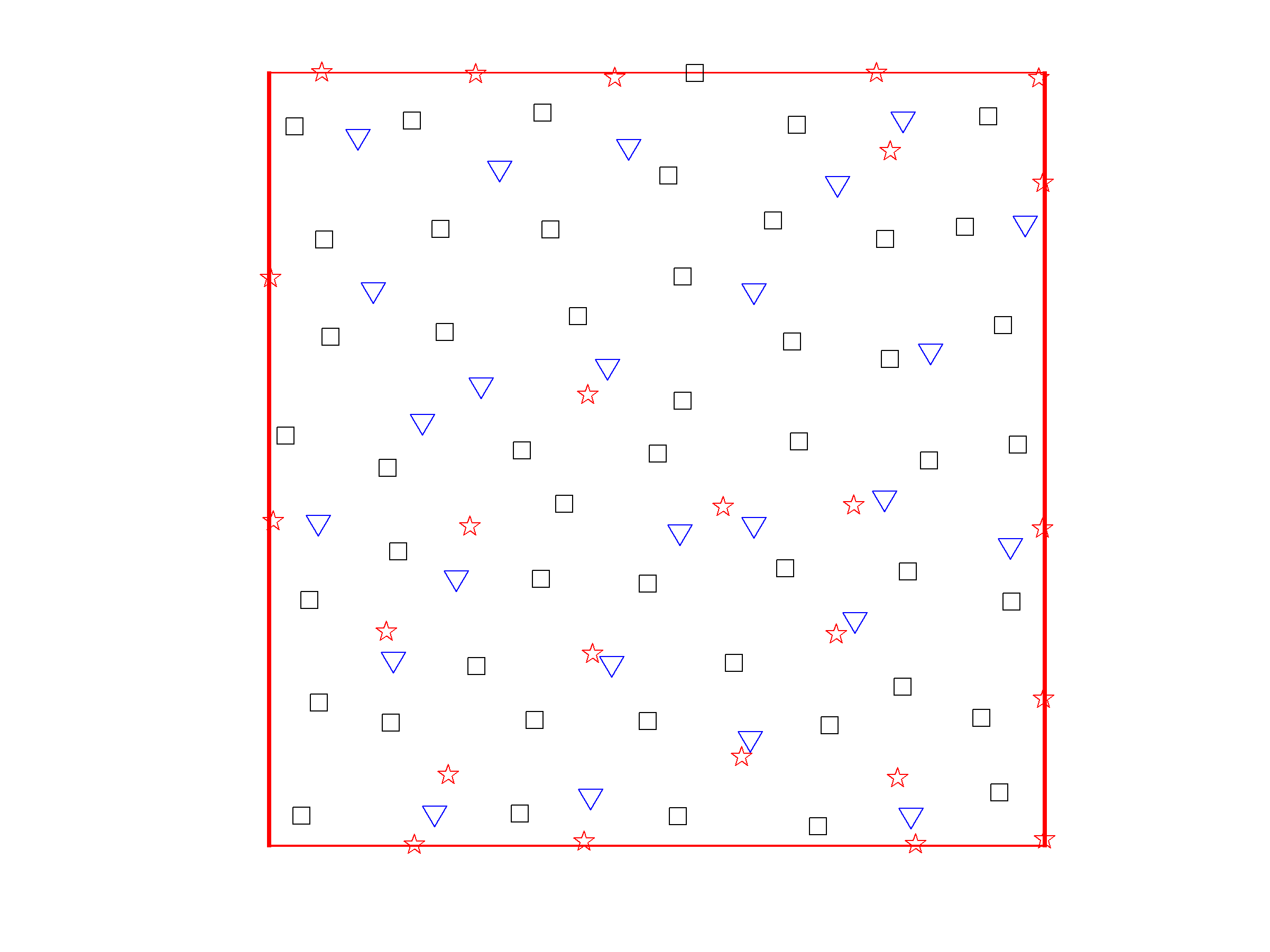} \includegraphics[width=.49\linewidth]{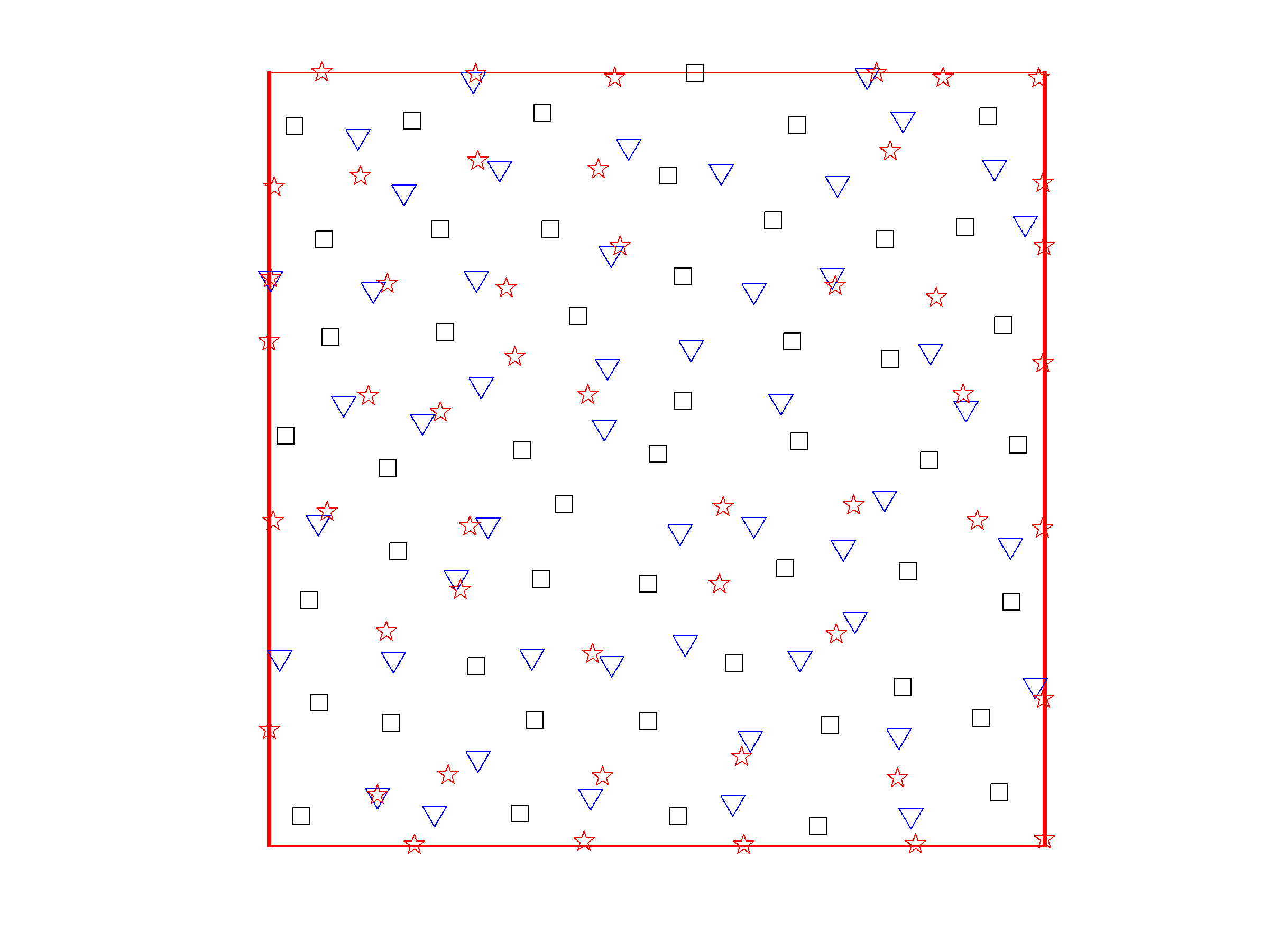}
\end{center}
\caption{\small Designs $\Xb_n=\KH(\emptyset,K,n)$ (black $\square$), $\KH(\Xb_n,K,m)$ (blue $\triangledown$) and  $\MN_2(\emptyset,\bK_{|n},m)$ (red $\bigstar$) for $n=50$ and $m=25$ (left), $m=50$ (right).}
\label{F:Design_Xn-AKH-BMN2_checkKn_d2}
\end{figure}

Figure~\ref{F:Design_Xn-AKH-BMN2_checkKn_d2} shows the design $\Xb_n=\KH(\emptyset,K,n)$ (black squares) and the validation designs $\KH(\Xb_n,K,m)$ (blue triangles) and  $\MN_2(\emptyset,\bK_{|n},m)$ (red stars) for $n=50$ and $m=25$ (left), and $n=50$, $m=50$ (right) when $d=2$ (note that $\KH(\Xb_n,K,25)\subset\KH(\Xb_n,K,50)$ and $\MN_2(\emptyset,\bK_{|n},25)\subset\MN_2(\emptyset,\bK_{|n},50)$). For the two values of $m$ considered, $\KH(\Xb_n,K,m)$ looks more evenly spread in $\SX$ than $\MN_2(\emptyset,\bK_{|n},m)$, even if both designs are well interlaced with $\Xb_n$.

The quantitative comparison below of the space-filling properties of the different designs considered relies on their covering and packing (or separating) radii, respectively defined by
\bea
\CR(\Xb_s) = \max_{\xb\in\SX} \min_{1\leq i \leq s} \|\xb-\xb_i\| \mbox{ and } \PR(\Xb_s) = \frac12\, \min_{i\neq j} \|\xb_i-\xb_j\|
\eea
when $\Xb_s=\{\xb_1,\ldots,\xb_s\}$. When $d\leq 4$, the exact value of $\CR(\Xb_s)$ is calculated by Vorono\"i tessellation \citep{P-JSFdS2017}; when $d>4$, we under-approximate $\CR(\Xb_s)$ by $\max_{\xb\in\SX_{Q'}} \min_{1\leq i \leq s} \|\xb-\xb_i\|$, with $\SX_{Q'}$ given by the first $2^{19}$ points of a scrambled Sobol' sequence complemented with a $3^d$ full factorial design (so that $Q'=2^{19}+3^d$).

\begin{figure}[ht!]
\begin{center}

\includegraphics[width=.45\linewidth]{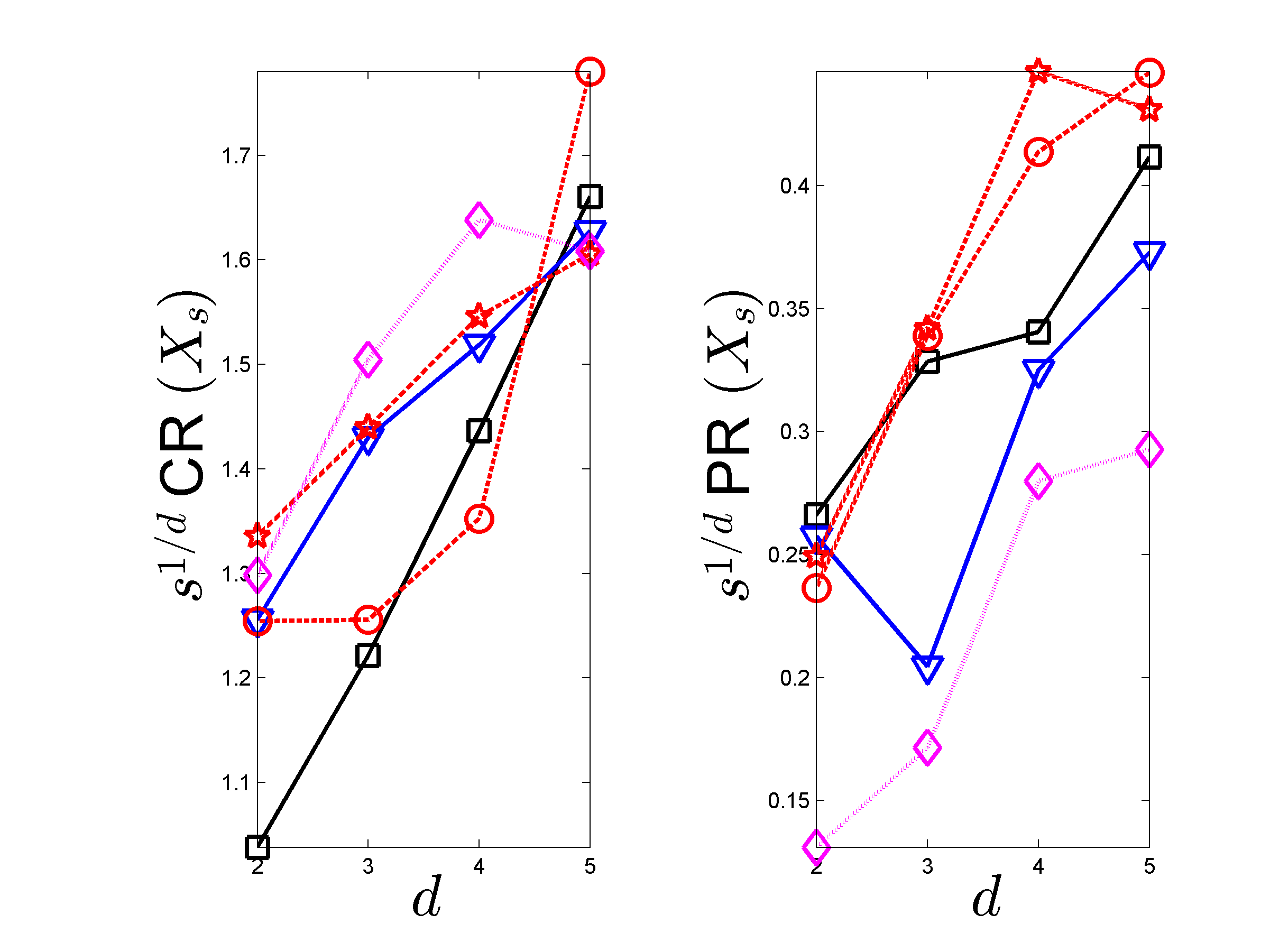}
\includegraphics[width=.45\linewidth]{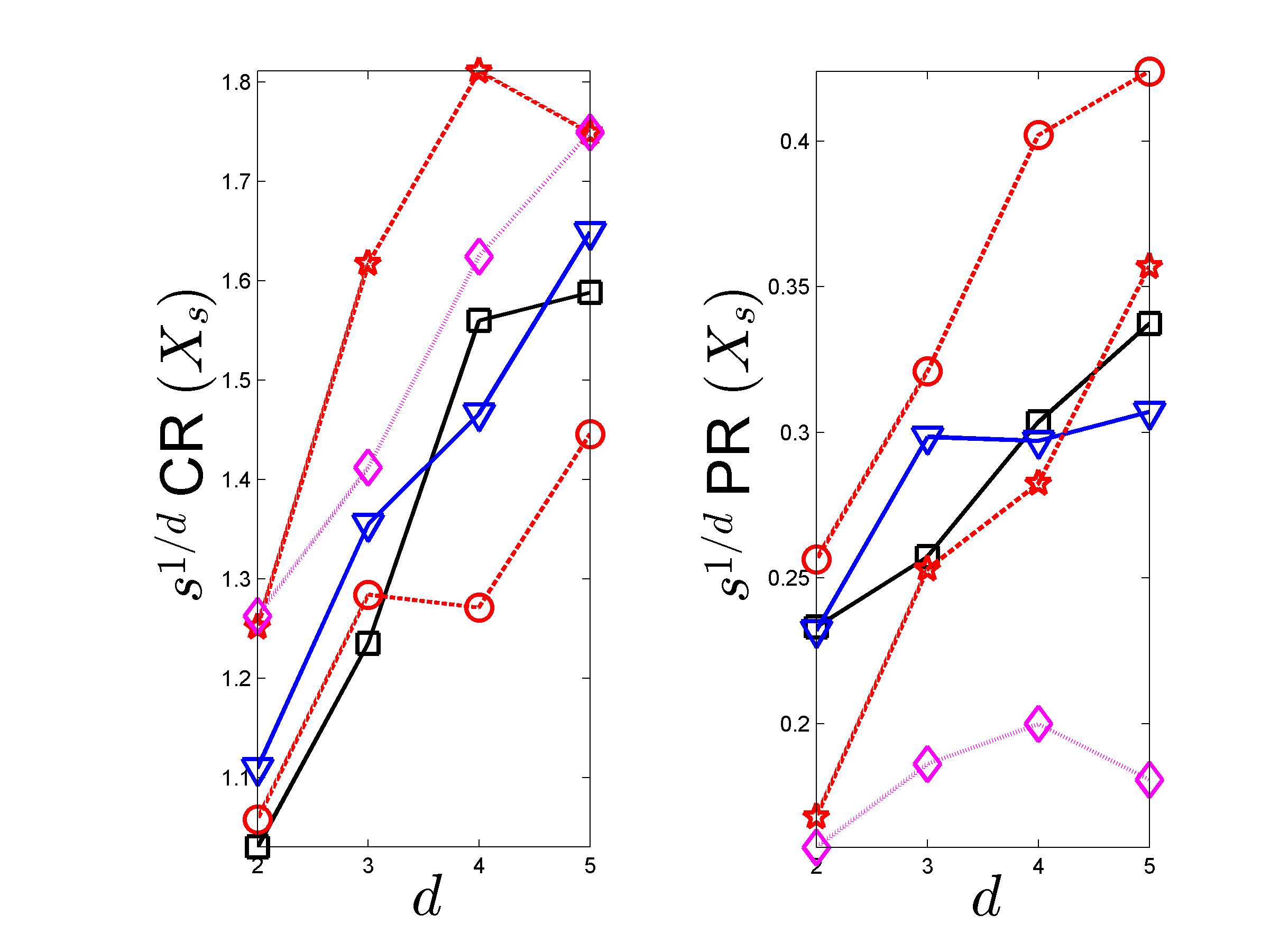}
\end{center}
\caption{\small Renormalized values of $\CR$ and $\PR$ for $\Xb_n=\KH(\emptyset,K,n)$ (black $\square$), $\KH(\Xb_n,K,m)$ (blue $\triangledown$), $\KH(\emptyset,\bK_{|n},k,m,\setminus\Xb_n)$, see (\textit{i}) in Section~\ref{S:kh-validation} (red $\circ$), and $\MN_2(\emptyset,\bK_{|n},m)$ (red $\bigstar$); first $m$ points $\Sb_m$ of a scrambled Sobol' sequence (magenta $\lozenge$).
Left column: $n=m=50$; right column: $n=200$, $m=100$.}
\label{F:CRPR_all_d2-5}
\end{figure}

Figure~\ref{F:CRPR_all_d2-5} presents the values of $\CR$ and $\PR$ (multiplied by $s^{1/d}$ for a design of size $s$) obtained for $\Xb_n=\KH(\emptyset,K,n)$ (black squares), $\KH(\Xb_n,K,m)$ (blue triangles), and $\KH(\emptyset,\bK_{|n},m)$ (red circles) and $\MN_2(\emptyset,\bK_{|n},m)$ (red stars), with $n=m=50$ on the left column and $n=200$, $m=100$ on the right. The magenta diamonds correspond to $\Sb_m$ given by the first $m$ points of a scrambled Sobol' sequence.

The designs constructed by kernel herding and its variants have good space-filling performance, typically better, and often much better, than Sobol' points $\Sb_m$. On the left column $m=n$, and we can directly compare the space-filling performance of $\Zb_m $ and $\Xb_n$. The good space-filling properties of $\Xb_n=\KH(\emptyset,K,n)$ tend to deteriorate when considering its continuation $\KH(\Xb_n,K,m)$. Some other constructions sometimes compare favorably to $\Xb_n$ in terms of $\CR$, or $\PR$, or even both. It is true in particular for $\KH(\emptyset,\bK_{|n},k,m,\setminus\Xb_n)$, see (\textit{i}) in Section~\ref{S:kh-validation}, for which we continue iterations until number of selected points not in $\Xb_n$ equals $m$: it is almost uniformly better than $\KH(\Xb_n,K,m)$. This opens interesting perspectives in terms of construction of space-filling designs.
$\MN_2(\emptyset,\bK_{|n},m)$ performs significantly worse; its rather poor space-filling properties were already apparent on Figure~\ref{F:Design_Xn-AKH-BMN2_checkKn_d2}.

\paragraph{b) IMSE and ISE performance.}

Below we compare the values of $\bDelta(\Zb_m,\Xb_n)$ given by \eqref{delta-MMD} for different designs $\Zb_m$. For all designs considered, weighted or not,
$\bDelta(\Zb_m,\Xb_n)$ is computed directly from the associated measure $\zeta_m$ as $\SE_{\bK_{|n}}^{1/2}(\zeta_m-\mu_Q)$, see \eqref{delta-MMD} and \eqref{MMD2-w}.

We also consider
\bea 
\Delta(\Zb_m,\Xb_n)= |\hIMSE(\Zb_m,\Xb_n)-\IMSE(\Xb_n)|\,,
\eea
where
\be\label{IMSE_n}
\hIMSE(\Zb_m,\Xb_n) &=& \frac{1}{m} \sum_{i=1}^m \Ex\left\{\left[\kb_n\TT(\zb_i)\yb_n-f(\zb_i)\right]^2\right\} = \frac{1}{m} \sum_{i=1}^m K_{|n}(\zb_i,\zb_i) 
\ee
and $\IMSE(\Xb_n)$ is given by \eqref{IMSE}, which is approximated by a discrete sum $\hIMSE(\SX_{Q''},\Xb_n)$, where $\SX_{Q''}$ corresponds to $Q''=2^{19}$ points of a scrambled Sobol' sequence. When some weights $\wb^{(m)}$ are associated with $\Zb_m$, with $\wb^{(m)}=\hat\wb(\Zb_m,C)$ or $\wb^{(m)}=\breve\wb(\Zb_m,C)$ for a kernel $C$, see \eqref{wopt-sum=1} and \eqref{newww}, we use
\bea
\hIMSE([\Zb_m,\wb^{(m)}],\Xb_n) = \sum_{i=1}^m \{\wb^{(m)}\}_i\, K_{|n}(\zb_i,\zb_i)
\eea
in $\Delta(\Zb_m,\Xb_n)$ instead of \eqref{IMSE_n}. The minimization of $\bDelta(\Zb_m,\Xb_n)$ is not equivalent to that of $\Delta(\Zb_m,\Xb_n)$, and we shall see that the designs constructed for the former are not necessarily the most efficient for the latter. Note that the evaluation of $\IMSE(\Xb_n)$ is much easier than that of $\ISE(\Xb_n)$; see, e.g., \citep{GP-SIAM_2014, GP-CSSC_2016, GP-CSDA2016} for the construction of designs $\Xb_n$ that minimize $\IMSE(\Xb_n)$.

\begin{figure}[ht!]
\begin{center}

\includegraphics[width=.45\linewidth]{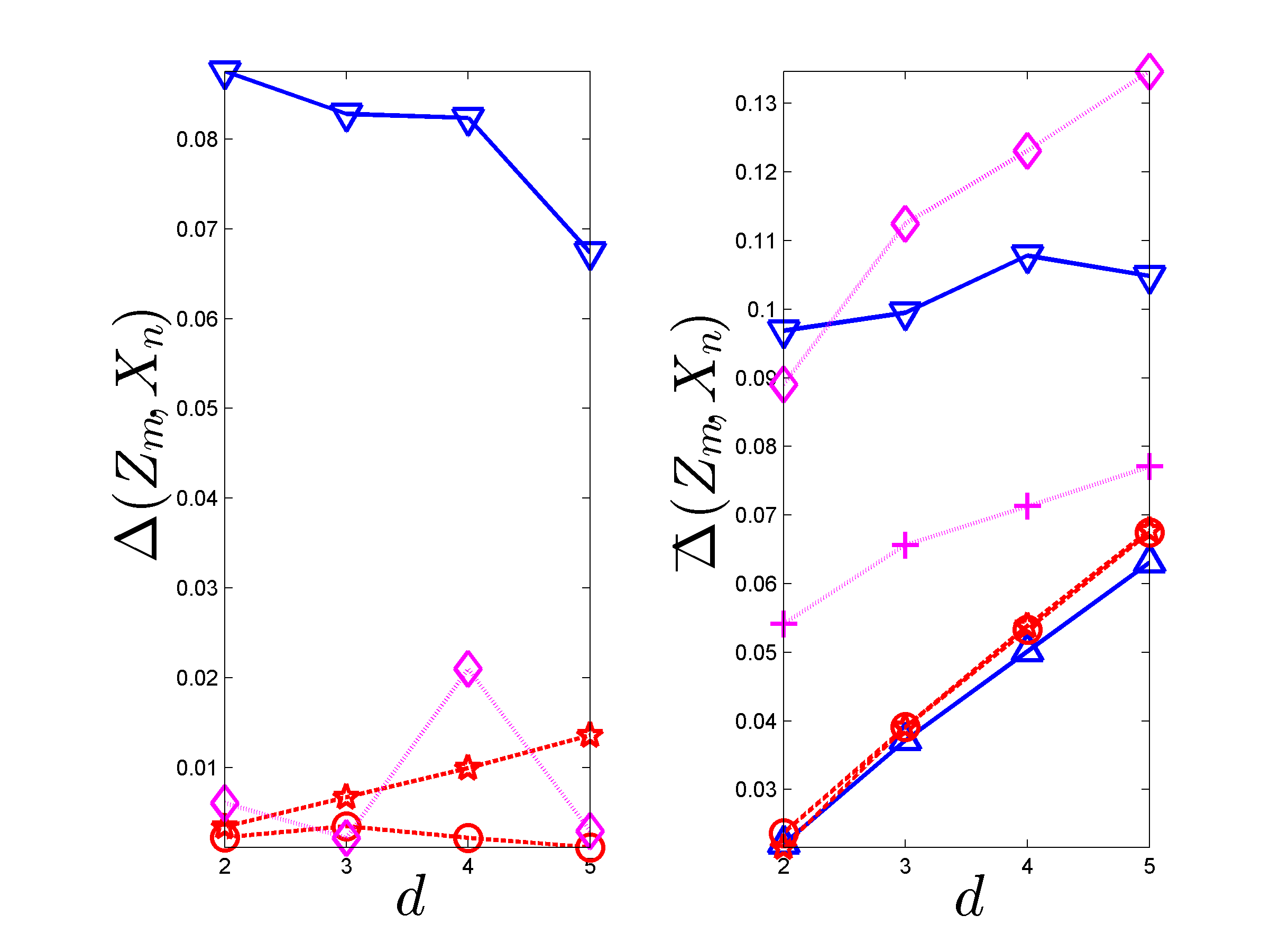}
\includegraphics[width=.45\linewidth]{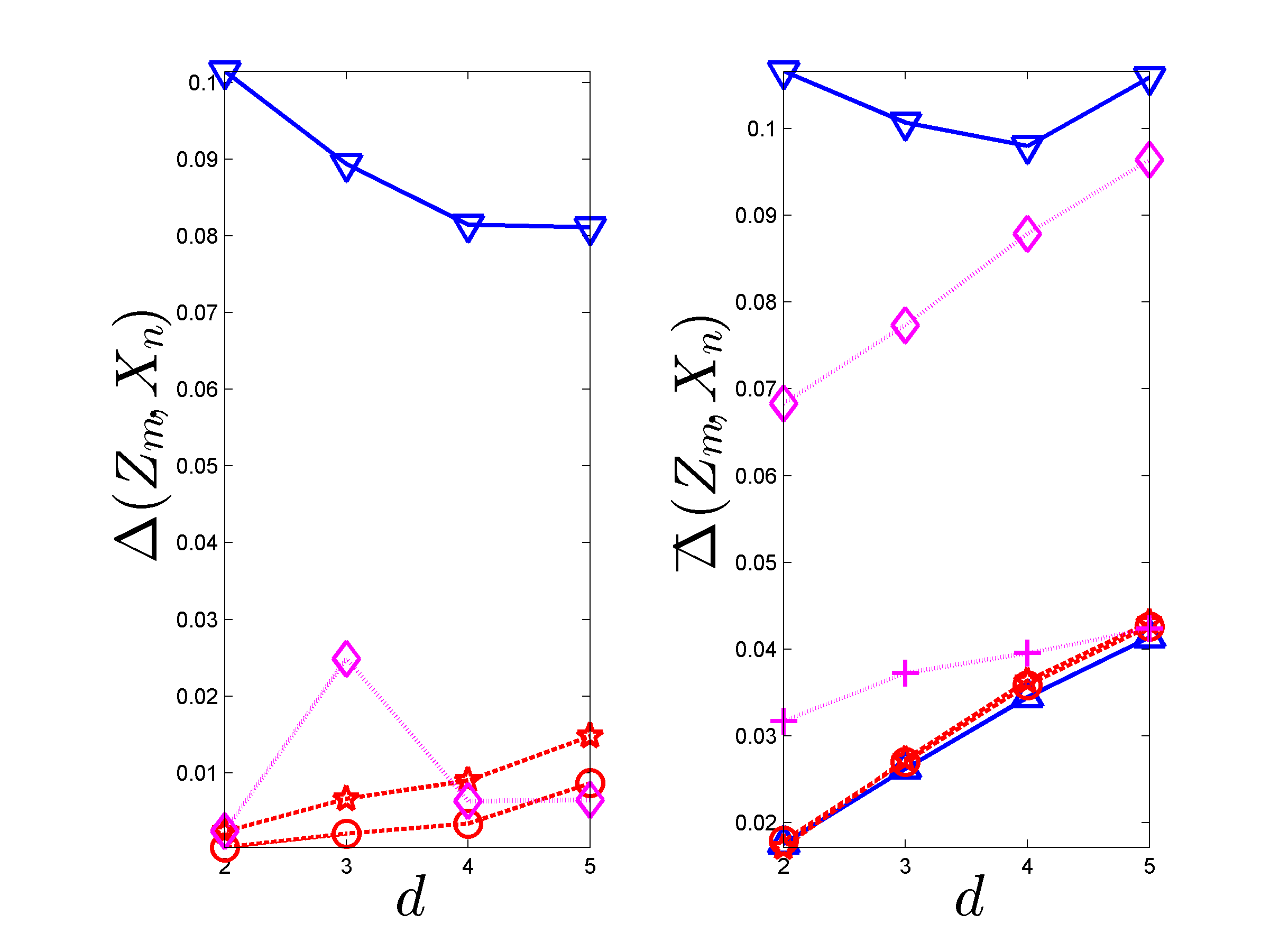}
\end{center}
\caption{\small $\Delta(\Zb_m,\Xb_n)$ and $\bDelta(\Zb_m,\Xb_n)$ for $\Zb_m=\KH(\Xb_n,K,m)$ (blue $\triangledown$) and $[\Zb_m,\breve\wb(\Zb_m,\bK_{|n})]$ (blue $\vartriangle$), $\KH(\emptyset,\bK_{|n},k,m,\setminus\Xb_n)$ with weights $1/k$ (red $\circ$) and $[\Zb^{''}_m,\breve\wb^{(m)}]=\MN_2(\emptyset,\bK_{|n},m)$ (red $\bigstar$);  first $m$ points $\Sb_m$ of a scrambled Sobol' sequence (magenta $\lozenge$), $[\Sb_m,\breve\wb(\Sb_m,\bK_{|n})]$ (magenta $+$). Left column: $n=m=50$; right column: $n=200$, $m=100$.}
\label{F:Delta_all_d2-5}
\end{figure}

Performances in terms of $\Delta(\Zb_m,\Xb_n)$ are shown on Figure~\ref{F:Delta_all_d2-5}, with Sobol' points $\Sb_m$ corresponding to magenta diamonds and $\Zb_m=\KH(\Xb_n,K,m)$ to blue triangles down. The designs $\KH(\emptyset,\bK_{|n},k,m,\setminus\Xb_n)$ correspond to red circles and $[\Zb^{''}_m,\breve\wb^{(m)}]=\MN_2(\emptyset,\bK_{|n},m)$ to red stars; $m=n=50$ on the left column, $n=200$ and $m=100$ on the right. In $\KH(\emptyset,\bK_{|n},k,m,\setminus\Xb_n)$, all points receive the same weight $1/k$ with $k>m$; see (\textit{i}) in Section~\ref{S:kh-validation}.
$\Delta(\Sb_m,\Xb_n)$ is much smaller than the values obtained for $\KH(\Xb_n,K,m)$. This could be anticipated from Figures~\ref{F:Pxi&Pmu_KH_Cn2_d1_n4_m4_M32}. It is related to the stronger variability of $K_{|n}(\zb_i,\zb_i)$ for Sobol' points, which are distributed independently of $\Xb_n$, than for the designs $\Zb_m$ constructed by kernel herding, which tend to fill the holes left by $\Xb_n$. For those designs, each $\zb_i$ is selected far away from its closest $\xb_j$, all $K_{|n}(\zb_i,\zb_i)$ tend to be large and $\hIMSE(\Zb_m,\Xb_n)$ tends to severely overestimate $\IMSE(\Xb_n)$.
The designs constructed with $\bK_{|n}$ compensate this effect by weight reduction and behave more similarly to Sobol' points: in $\KH(\emptyset,\bK_{|n},k,m,\setminus\Xb_n)$ all points receive the same weight $1/k<1/m$; in $[\Zb^{''}_m,\breve\wb^{(m)}]=\MN_2(\emptyset,\bK_{|n},m)$ the total mass is smaller than one.

Consider now our criterion of interest, $\bDelta(\Zb_m,\Xb_n)$. The same symbols as above are used to represent the different designs, but two more designs are considered: $[\Sb_m,\breve\wb(\Sb_m,\bK_{|n})]$ with magenta plus and $[\KH(\Xb_n,K,m),\breve\wb(\KH(\Xb_n,K,m),\bK_{|n})]$ with blue triangles up. We can see that the introduction of weights $\breve\wb(\Zb_m)$ has a major effect on the reduction of $\bDelta(\Zb_m,\Xb_n)$; $\KH(\emptyset,\bK_{|n},k,m,\setminus\Xb_n)$ and $[\Zb^{''}_m,\breve\wb^{(m)}]=\MN_2(\emptyset,\bK_{|n},m)$ have very good performance too.

\begin{figure}[ht!]
\begin{center}

\includegraphics[width=.35\linewidth]{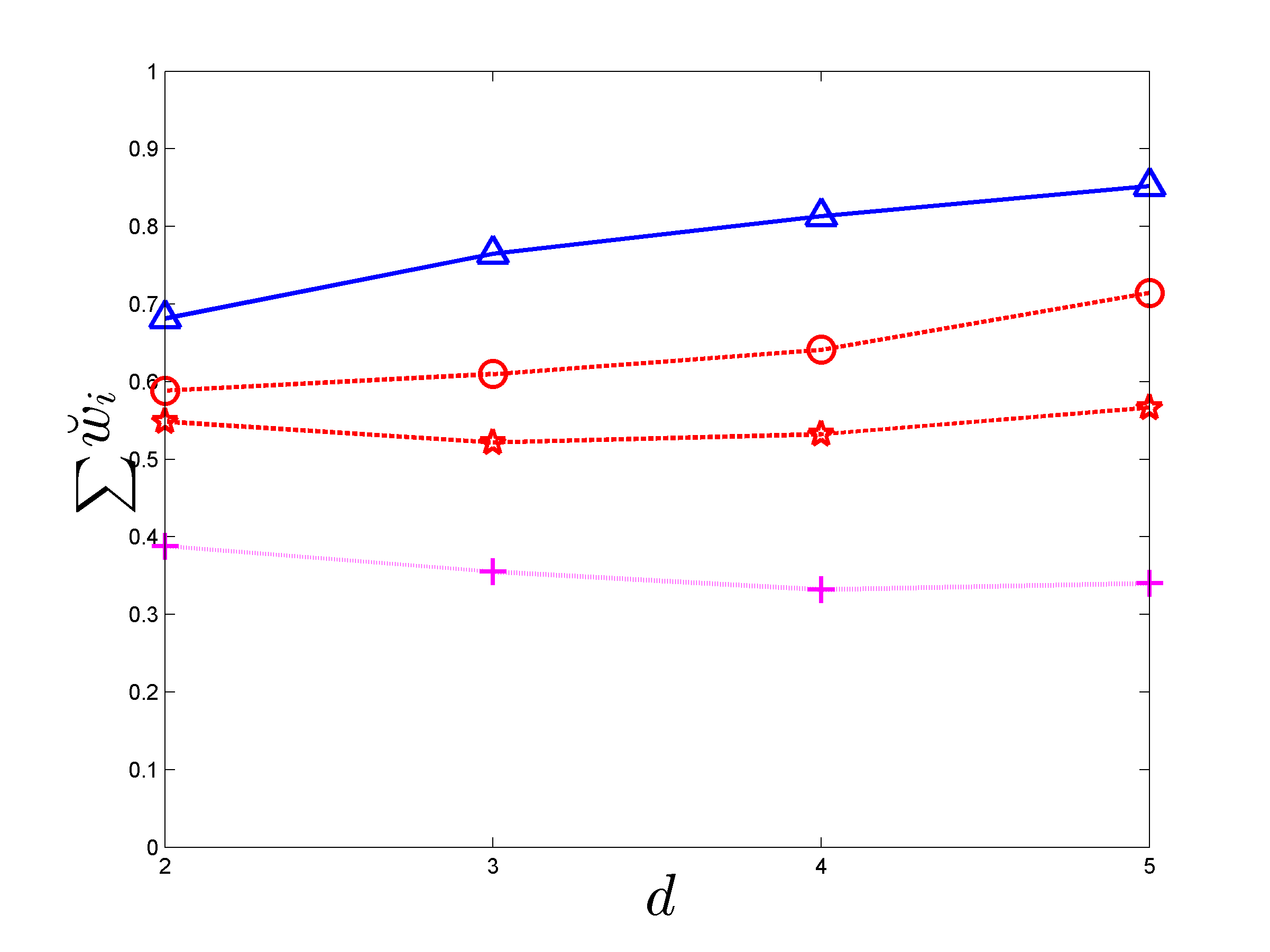}
\includegraphics[width=.35\linewidth]{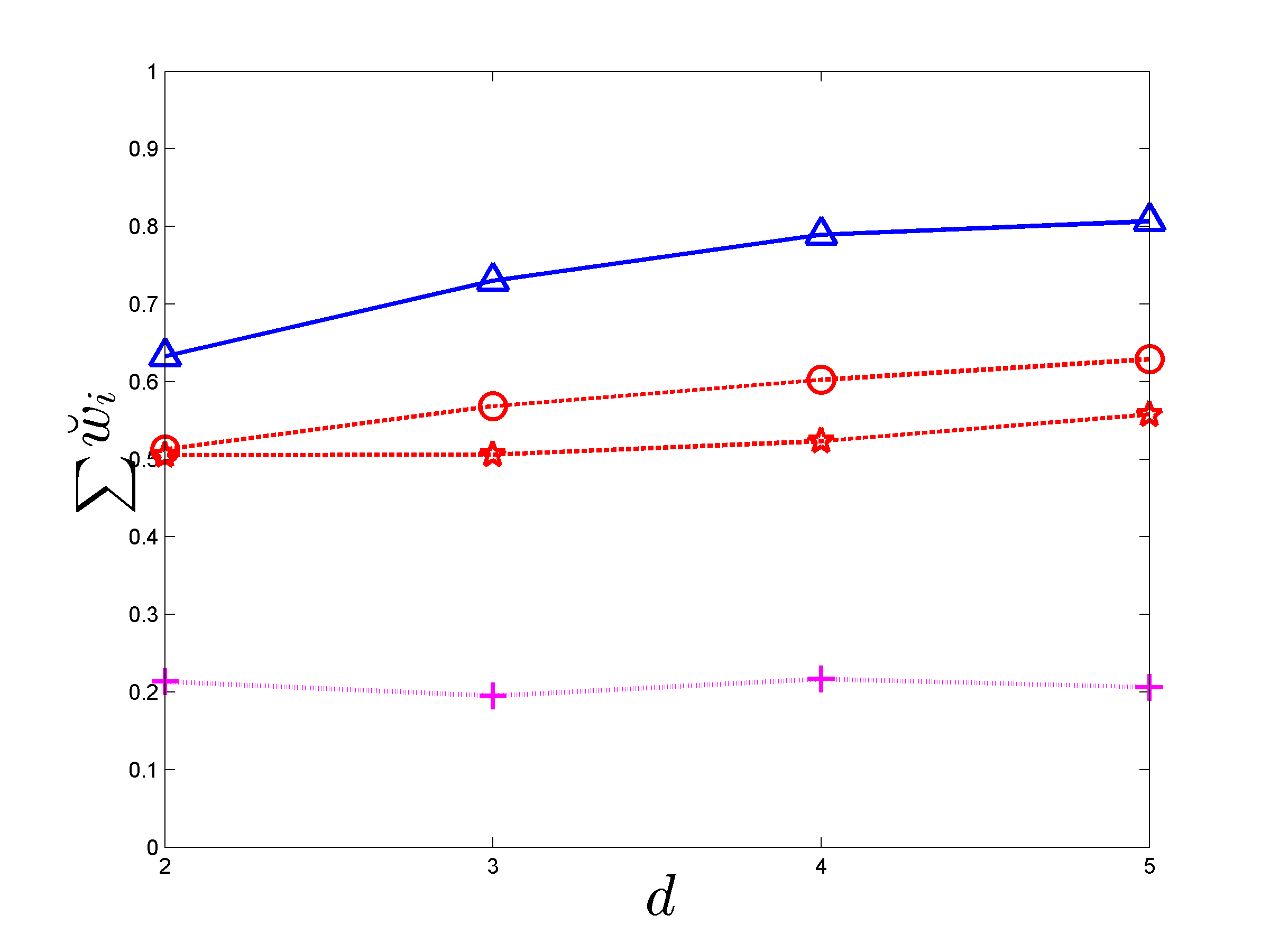}
\end{center}
\caption{\small $\sum_{i=1}^m \breve w_i$ for $[\Zb_m,\breve\wb(\Zb_m,\bK_{|n})]$ (blue $\vartriangle$),  $[\Zb^{''}_m,\breve\wb^{(m)}]=\MN_2(\emptyset,\bK_{|n},m)$ (red $\bigstar$) and $[\Sb_m,\breve\wb(\Sb_m,\bK_{|n})]$ (magenta $+$), with $\Sb_m$ given by the first $m$ points of a scrambled Sobol' sequence. For $\KH(\emptyset,\bK_{|n},k,m,\setminus\Xb_n)$ (red $\circ$), the total mass equals $m/k$. Left column: $n=m=50$; right column: $n=200$, $m=100$.}
\label{F:Sumw_all_d2-5}
\end{figure}

Figure~\ref{F:Sumw_all_d2-5} shows the total mass $\sum_{i=1}^m \breve w_i$ for the designs $[\Sb_m,\breve\wb(\Sb_m,\bK_{|n})]$ (magenta plus)
$[\Zb_m,\breve\wb(\Zb_m,\bK_{|n})]$ (blue triangles up), $[\Zb^{''}_m,\breve\wb^{(m)}]=\MN_2(\emptyset,\bK_{|n},m)$ (red stars) and $m/k$ for the design $\KH(\emptyset,\bK_{|n},k,m,\setminus\Xb_n)$ (red circles); $m=n=50$ on the left column, $n=200$ and $m=100$ on the right. There is no strict relation between total mass and performance in terms of $\bDelta(\Zb_m,\Xb_n)$ shown on Figure~\ref{F:Delta_all_d2-5}, indicating that it is the interplay between the location of the points and the weighing that matters. Note in particular that $[\Zb_m,\breve\wb(\Zb_m,\bK_{|n})]$ and $[\Zb^{''}_m,\breve\wb^{(m)}]=\MN_2(\emptyset,\bK_{|n},m)$ have quite different weighings although they have similar values of $\bDelta(\Zb_m,\Xb_n)$ on Figure~\ref{F:Delta_all_d2-5}.

\section{Examples of validation design for ISE estimation}\label{S:numerical-results}

\subsection{Separable kernels}\label{S:separableK}


The substitution of a finite set $\SX_Q$ for $\SX$ and of the uniform measure on $\SX_Q$ for $\mu$ yields a drastic simplification of calculations in the evaluation of the MMD $\mg_{\bK_{|n}}(\zeta_m,\mu)$ and in the algorithmic construction of designs by kernel herding and its variants. However, for large $d$ we need to take $Q$ very large to make $\SX_Q$ dense enough in $\SX$, and another approach is required if we want to maintain a reasonable accuracy.


A bottleneck in the application of kernel herding is the need to calculate $P_{\bK_{|n},\mu}(\zb)$ for many $\zb$ in order to choose $\zb_{k+1}$ in \eqref{z-n+1}. An additional difficulty for the evaluation of $\mg_{\bK_{|n}}(\zeta_m,\mu)$ is the need to compute $\SE_{\bK_{|n}}(\mu)$, see \eqref{MMD2-w}. However, when $K$ is a separable (tensor-product) kernel, both $P_{\bK_{|n},\mu}$ and $\SE_{\bK_{|n}}(\mu)$ can be calculated explicitly.

Since $\mu$ is uniform on $\SX=[0,1]^2$, we can write $\mu(\dd\xb)=\prod_{i=1}^d \mu_1(\dd x_i)$ with $\mu_1$ the uniform measure on $[0,1]$ and $\xb=(x_1,\ldots,x_d)\TT$. For a separable (or tensor-product) kernel $K$, such that
\bea
K(\xb,\xb')=\prod_{i=1}^d K_i(x_i,x'_i)\,,
\eea
where $\xb=(x_1,\ldots,x_d)\TT$ and $\xb'=(x'_1,\ldots,x'_d)\TT$, we have
\bea
\SE_{K}(\mu) = \prod_{i=1}^d \SE_{K_i}(\mu_1) \mbox{ and }
P_{K,\mu}(\xb) = \prod_{i=1}^d \int_{\SX_i} K_i(x_i,x'_i)\, \mu_1(\dd x'_i) = \prod_{i=1}^d P_{K_i,\mu_1}(x_i) \,.
\eea
One may refer to \citet{SzaboS2018} for connections between positive-definiteness properties of the $K_i$ and those of $K$.
The expressions of $\SE_{K_i}(\mu_1)$ and $P_{K_i,\mu_1}(\cdot)$ are available for many kernels $K_i$; see \citet{PZ2020-SIAM} and the references therein.

\vsp
Before deriving the expressions of $P_{\bK_{|n},\mu}(\xb)$ and $\SE_{\bK_{|n}}(\mu)$, we introduce some notation.
Denote by $\bmOb_{K,n}$ and $\bmGb_{K,n}$ the $n\times n$ matrices with respective elements
\bea
\{\bmOb_{K,n}\}_{j,k}= \prod_{i=1}^d \beta_{K_i}({x_j}_i,{x_k}_i) \mbox{ and }  \{\bmGb_{K,n}\}_{j,k}= \prod_{i=1}^d \mg_{K_i}({x_j}_i,{x_k}_i) \,,
\eea
and by $\bmob_{K,n}(\xb)$ the vector with $j$-th component
\bea
\{\bmob_{K,n}(\xb)\}_j= \prod_{i=1}^d \beta_{K_i}({x_j}_i,x_i)\,,
\eea
where ${x_j}_i$ (respectively, ${x_k}_i$) is the $i$-th component of $\xb_j$ (respectively, $\xb_k$), and
\bea
\beta_{K_i}(r,s) &=& \int_\SX K_i(r,t)K_i(s,t)\, \mu_1(\dd t) \,, \ i=1,\ldots,d\,,\\
\mg_{K_i}(r,s) &=& \int_{\SX^2} K_i(r,t)K_i(s,u)K_i(t,u)\, \mu_1(\dd t)\mu_1(\dd u) \,, \ i=1,\ldots,d\,.
\eea
Then, using \eqref{bK_|n}, direct calculation gives
\bea
P_{\bK_{|n},\mu}(\xb) &=& 2\,P_{K^2,\mu}(\xb)- 4\,\kb_n\TT(\xb)\Kb_n^{-1}\bmob_{K,n}(\xb) + 2\, \kb_n\TT(\xb)\Kb_n^{-1}\bmOb_{K,n}\Kb_n^{-1}\kb_n(\xb)  \\
&& + \left[1-\kb_n\TT(\xb)\Kb_n^{-1}\kb_n(\xb)\right]\left[1-\tr(\Kb_n^{-1}\bmOb_{K,n})\right] \,, \\
\SE_{\bK_{|n}}(\mu) &=& 2\,\SE_{K^2}(\mu) - 4\, \tr(\Kb_n^{-1}\bmGb_{K,n}) + 2\, \tr\left[(\Kb_n^{-1}\bmGb_{K,n})^2\right]
+\left[1-\tr(\Kb_n^{-1}\bmOb_{K,n})\right]^2\,.
\eea


The expressions of $P_{K^2,\mu_1}(x)$, $\SE_{K^2}(\mu_1)$, $\beta_{K}(u,v)$ and $\mg_{K}(u,v)$, $x,u,v\in[0,1]$, for $\mu_1$ uniform on $[0,1]$ and $K_i(x,x')$ a Mat\'ern~3/2 kernel \eqref{K32} are given in Appendix~B, making the expressions of $P_{\bK_{|n},\mu}(\xb)$ and $\SE_{\bK_{|n}}(\mu)$ available in closed form when $K(\xb,\xb')$ is the product of uni-dimensional Mat\'ern~3/2 kernels and $\mu$ is uniform on $\SX=[0,1]^d$. Similar calculations can be conducted for other kernels.

\subsection{Numerical results}\label{S:Numerical results}

We use test functions given by random multivariate polynomials in dimension $d=2,\ldots,10$, with $n=100$ and $m=50$, generated as indicated in Appendix~C, the set $\SSL$ in \eqref{Poly} being constrained by $N=n/2$, $p=7$ and $p_T=25$. We take $\ma=1/2$ in \eqref{Qrot}, $\ml_i=1/[(i+1)^2\,\tau^i]$, where $\tau=\max_{i=1,\ldots,d} \sum_{j=1}^d |\{\Qb\}_{i,j}|$ (the renormalization by $\tau^i$ accounts for the fact that points $\{\Qb(\xb-\1b_d/2)+\1b_d/2\}_i$ do not belong to $[0,1]$).

For each $d=2,\ldots,10$, we generate $r=100$ random functions $f^{(j)}$, $j=1,\ldots,r$. For each $f^{(j)}$, $\Xb_n$ corresponds to the first $n$ points of a scrambled Sobol' sequence, the next $m$ points of the sequence are denoted $\Sb_m$ and form one of the validation designs considered in the comparison. The second design considered is $\Zb_m=\KH(\Xb_n,K,m)$, constructed by kernel herding with a candidate set $\SX_Q$ given by the first $Q=2^{16}$ points of another scrambled Sobol' sequence. We also consider random designs $\Rb_m$ made of $m$ points independently uniformly distributed in $[0,1]^d$. A different design $\Xb_n$, $\Sb_m$, $\Zb_m$, $\Rb_m$ and candidate set $\SX_Q$ is used for each random $f^{(j)}$ generated, but we omit the index $j$ in the notation.
The kernel $K$ is the tensor product of univariate Mat\'ern 3/2 kernels \eqref{K32}.  The construction of $\KH(\Xb_n,K,m)$ by kernel herding and the computation of the weights $\breve\wb_m$ given by \eqref{newww} exploit the results of Section~\ref{S:separableK}.

We set $\mt=n^{1/d}$ in \eqref{K32} to construct $\KH(\Xb_n,K,m)$ (it is the space-filling property of $\Zb_m$ that matters here), but to estimate $\ISE(\Xb_n)$ we use $\mt=\mt_n^{(j)}$ estimated by Leave-One-Out Cross Validation (LOO CV) applied to the centered data $\widetilde\yb_n^{(j)}=\yb_n^{(j)}-\bar y_n^{(j)}\1b_n$, with $\bar y_n^{(j)}=\1b_n\TT\yb_n^{(j)}/n$ the empirical mean of $\yb_n^{(j)}=(f^{(j)}(\xb_1),\ldots,f^{(j)}(\xb_n))$. Following \citet{Dubrule83}, $\mt_n^{(j)}$ minimizes
\be\label{ISE-LOO}
\widehat\ISE_{LOO}{(j)} = \frac{1}{n}\, \sum_{i=1}^n [f^{(j)}(\xb_i)-\eta_{n,-i}^{(j)}(\xb_i)]^2
= \frac{1}{n}\, (\widetilde\yb_n^{(j)})\TT\Kb_n^{-1}\Db_n\Kb_n^{-1}\widetilde\yb_n^{(j)}
\ee
with respect to $\mt\in\mathds{R}^+$,
where $\eta_{n,-i}^{(j)}(\xb)$ uses the $n-1$ points in $\Xb_n\setminus\{\xb_i\}$ and $\Db_n$ is the diagonal matrix with elements $\{\Db_n\}_{i,i}=\{\Kb_n^{-1}\}_{i,i}^{-2}$; $\Kb_n$ depends on $\mt$ through \eqref{K32}.

The exact value of $\ISE^{(j)}=\ISE^{(j)}(\Xb_n)$ given by \eqref{ISE} is approximated by a discrete sum, with the uniform measure on $\SX_Q$ substituted for $\mu$. For each one of the designs $\Sb_m$, $\Zb_m$ and $\Rb_m$ we compute the optimum weights $\breve\wb$ for the kernel $\bK_{|n}$, and for each $f^{(j)}$ we compute
\bea
\widehat\ISE^{(j)}(\Zb_m,\Xb_n)= \frac{1}{m} \sum_{i=1}^m [f^{(j)}(\zb_i)-\eta_n^{(j)}(\zb_i)]^2 \,, \
\widehat\ISE^{(j)}(\Zb_m,\breve\wb_m,\Xb_n)= \sum_{i=1}^m \breve w_i [f^{(j)}(\zb_i)-\eta_n^{(j)}(\zb_i)]^2\,,
\eea
for the unweighted and weighted design, respectively, where $\eta_n^{(j)}(\xb)=\kb_n\TT(\xb)\Kb_n^{-1}\widetilde\yb_n^{(j)}+\bar y_n^{(j)}\1b_n$ and $\mt=\mt_n^{(j)}$ in $\kb_n$ and $\Kb_n$.
For each function $f^{(j)}$ and each design, weighted or not, we denote by $\rho^{(j)}$ the relative error.
\bea
\rho^{(j)} = \frac{\widehat\ISE^{(j)} - \ISE^{(j)}}{\ISE^{(j)} }
\eea

\begin{figure}[ht!]
\begin{center}
\includegraphics[width=.45\linewidth]{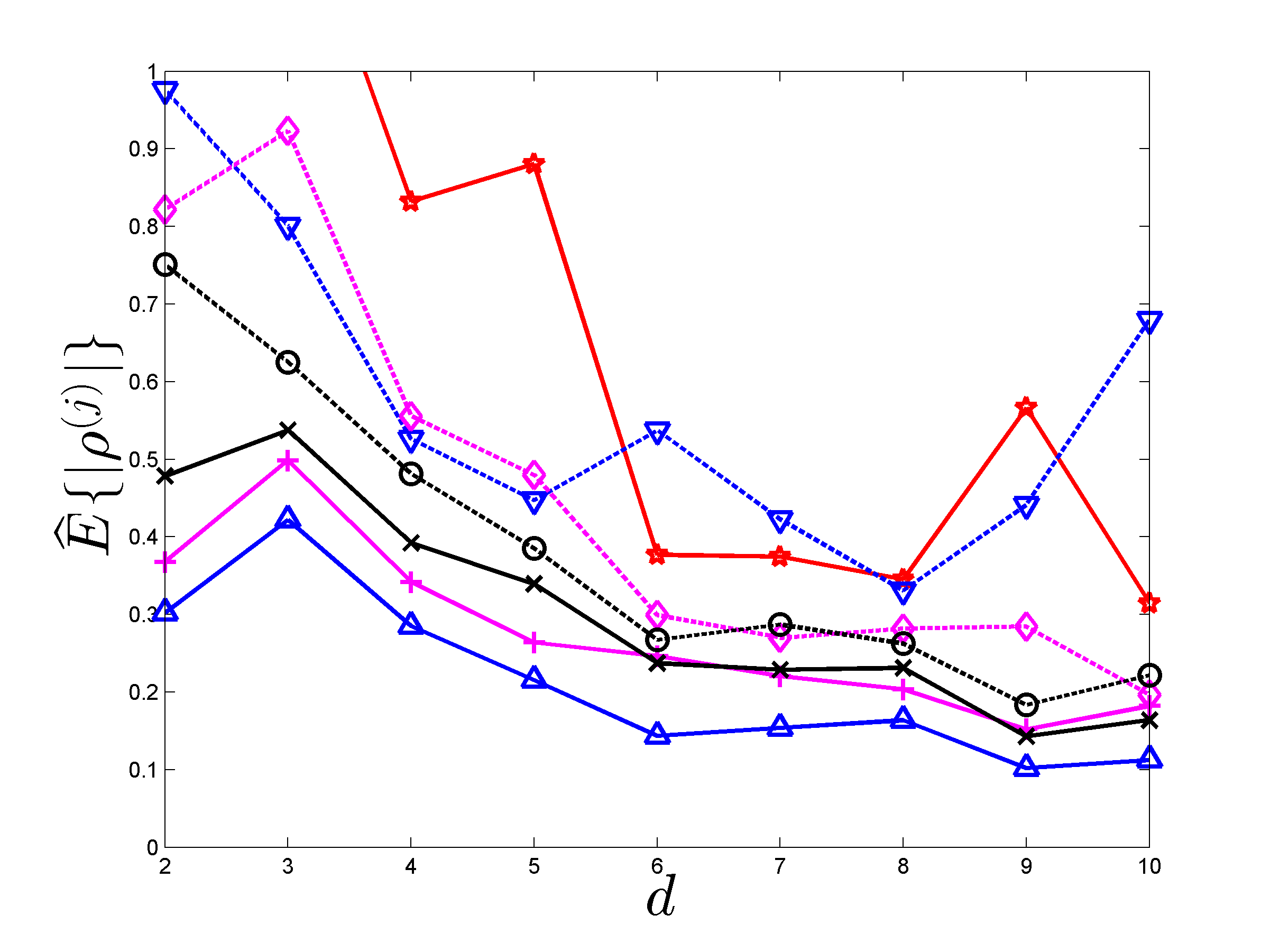}
\includegraphics[width=.45\linewidth]{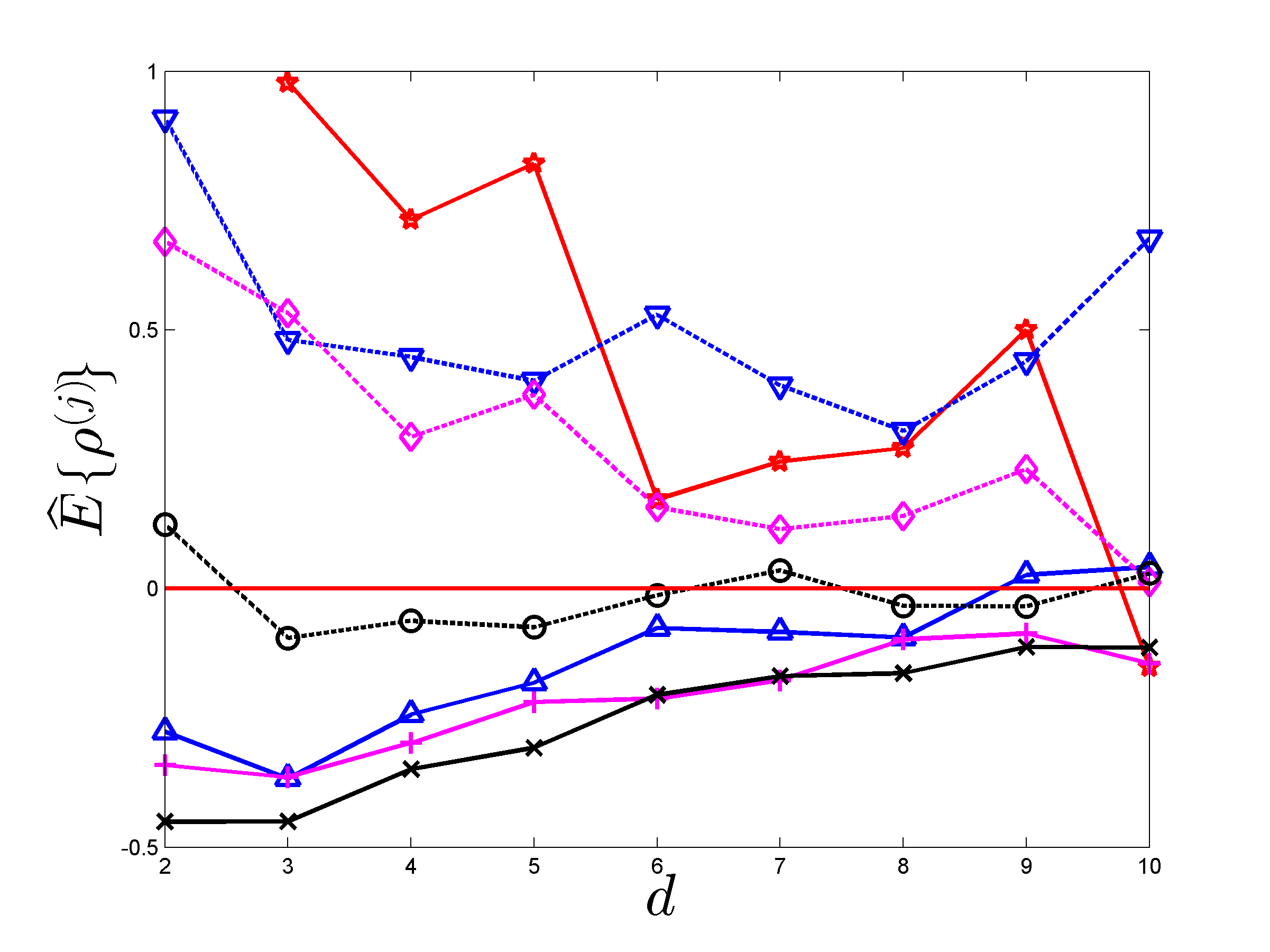}

\end{center}
\caption{\small Left: $\widehat\Ex\{|\rho^{(j)}|\}$; Right: $\widehat\Ex\{\rho^{(j)}\}$; for $d=2,\ldots,10$ and the designs $\Sb_m$ (magenta $\lozenge$), $[\Sb_m,\breve\wb_m(\Sb_m,\bK_{|n})]$ (magenta +), $\Zb_m=\KH(\Xb_n,K,m)$ (blue $\triangledown$), $[\Zb_m,\breve\wb_m(\Zb_m,\bK_{|n})]$ (blue $\vartriangle$), $\Rb_m$ (black $\circ$), $[\Rb_m,\breve\wb_m(\Rb_m,\bK_{|n})]$ (black $\times$); LOO CV (red $\bigstar$); $100$ repetitions, $n=100$, $m=50$.}
\label{F:Relative_ISE_error}
\end{figure}

The left panel of Figure~\ref{F:Relative_ISE_error} presents the empirical means $\widehat\Ex\{|\rho^{(j)}|\}=(1/r)\,\sum_{j=1}^r |\rho^{(j)}|$ as functions of $d=2,\ldots,10$ obtained for the different designs considered and for $\widehat\ISE_{LOO}{(j)}$ given by \eqref{ISE-LOO}; the right panel shows $\widehat\Ex\{\rho^{(j)}\}= (1/r)\,\sum_{j=1}^r \rho^{(j)}$. Unsurprisingly, LOO CV strongly overestimates $\ISE(\Xb_n)$ since (\textit{i}) each of the $n$ predictions in the summation in \eqref{ISE-LOO} uses $n-1$ design points only, and (\textit{ii}) each $\xb_i$ is far from the $n-1$ other design points. The superiority of the weighted design $[\Zb_m,\breve\wb_m(\Zb_m,\bK_{|n})]$ over the other ones is clear on the left panel; in particular weight reduction by $\breve\wb_m(\Zb_m,\bK_{|n})$ greatly improves the precision of ISE estimation, compare the two curves with triangles. Sobol' and random points behave similarly, with slightly better performance for Sobol' points in the weighted versions. The right panel provides information on the bias on ISE estimation. The three weighted designs (solid lines) underestimate $\ISE(\Xb_n)$. The unweighted random design $\Rb_m$ (dashed line with circles) has a small bias, but is of limited interest due to its large variability, as shown on the left panel. Both $\Sb_m$ and $\Zb_m$ tend to fill the holes left by $\Xb_n$ and therefore overestimate $\ISE(\Xb_n)$. Other designs, in particular based on the kernel $\bK_{|n}$ have also been considered, but they perform worse than $[\Zb_m,\breve\wb_m(\Zb_m,\bK_{|n})]$ and the results are not shown.

%

\section{Conclusions}\label{S:conclusions}

%
%
%
%
%
%

The construction of a validation design $\Zb_m$ aimed at estimating $\ISE(\Xb_n)$ for a given $\Xb_n$ can be casted as the choice of a design minimizing a maximum mean discrepancy for a particular kernel, conditional of $\Xb_n$. A sequence of nested validation designs can be obtained by incremental construction via kernel herding. Numerical experiments indicate that the most important characteristics of a good validation design are its space-filling properties (it should populate the holes left by $\Xb_n$ to properly explore the design space) and the weighting of its points (since evaluations far from the design points tend to overestimate the global error). 
What one would expect is that some combination of both is needed: if the validation points would sample the error well, no
weighting would be needed; if the points were space-filling, and the design very regular, a contant weight smaller than one would be almost optimal. In fact these factors play in antagonistic directions and some compromise is needed. 
A dedicated weighting method, based on a particular kernel, conditional on $\Xb_n$, has been proposed. Numerical simulations with random functions show the effectiveness of this weight reduction when it is applied to random or usual low-discrepancy designs. Performances are still better when the weight reduction is associated with a space-filling design that minimizes a kernel discrepancy: they are significantly better than those obtained with leave-one-out cross validation, which strongly overestimates $\ISE(\Xb_n)$.

\vspace{0.5cm}
\appendix
\vsp
\noindent{\Large\bf Appendix A: Characteristic kernels}
\vsp

A characteristic kernel $C$ defines a metric on the set of probability measures on $\SX$. 
It is called \emph{Integrally Strictly Positive Definite} (ISPD) when $\SE_C(\nu)>0$ for any nonzero signed measure $\nu$ on $\SX$, see \eqref{energy}, and \emph{Conditionally Integrally Strictly Positive Definite} (CISPD) when $\SE_C(\nu)>0$ for all nonzero signed measures $\nu$ on $\SX$ with total mass $\nu(\SX)=0$. An ISPD kernel is CISPD; a bounded ISPD kernel is SPD and defines an RKHS. When $C$ is uniformly bounded, it is characteristic if and only if it is CISPD; see \cite[Lemma~8]{SriperumbudurGFSL2010}. For instance, the isotropic squared exponential, Mat\'ern and generalized multiquadric kernels are ISPD.

The kernel $\bK_{|n}$ considered in this paper is positive definite but not strictly positive definite (and thus not ISPD). Indeed, $K_{|n}(\xb,\xb_i)=\bK_{|n}(\xb,\xb_i)=0$ for all $\xb$ and all $\xb_i$, $i=1,\ldots,n$, implying that $\mg_{\bK_{|n}}(\zeta,\xi)=0$ for any measures $\zeta$ and $\xi$ supported on $\Xb_n$.
Since $\mu$ is uniform and not supported on $\Xb_n$, it nevertheless makes sense to minimize $\mg_{\bK_{|n}}(\zeta_m,\mu)$.
The investigation of conditions under which $\mg_{\bK_{|n}}(\zeta,\mu)=0$ would imply $\zeta=\mu$, exploiting for instance the notion of universal kernel \citep{SriperumbudurFL2011}, is beyond the scope of this paper and we simply mention the following two points, concerning respectively $K_{|n}$ and $\bK_{|n}$. 

\begin{description}
  \item[(\textit{i})] Suppose that $K$ is ISPD. For any signed measure $\xi$ on $\SX$ and $\wb\in\mathds{R}^n$, define $\xi^{[\wb]}=\xi+\sum_{i=1}^n w_i\,\delta_{\xb_i}$. Then, using the notation of Section~\ref{S:kh-basics}, $\SE_K(\xi^{[\wb]}-\mu)=\SE_K(\xi-\mu)+\wb\TT\Kb_n\wb+2\,\wb\TT\pb_{K,n}(\xi-\mu) \geq 0$, with equality if and only if $\xi^{[\wb]}=\mu$ since $K$ is ISPD. Direct calculation gives
$\min_{\wb} \SE_K(\xi^{[\wb]}-\mu) = \SE_{K_{|n}}(\xi-\mu)$, and therefore $\SE_{K_{|n}}(\xi-\mu)\geq 0$ with equality if and only if $\xi^{[\wb]}=\mu$. As $\mu$ has no discrete component, $\xi^{[\wb]}=\mu$ implies $\wb=\0b$, and we get $\xi=\mu$.

    \item[(\textit{ii})]
Suppose now that $K$ is ISPD and continuous on $\SX$ and consider its Mercer decomposition: $K(\xb,\xb')=\sum_{i\geq 1} \ml_i \,\varphi_i(\xb)\varphi_i(\xb')$, $\ml_i>0$. It yields the following decomposition for $K^2$: $K^2(\xb,\xb')=\sum_{i,j\geq 1} \ml_i\ml_j\, \varphi_i(\xb)\varphi_j(\xb)\varphi_i(\xb')\varphi_j(\xb')$, and $\SE_{K^2}(\nu)=0$ for some signed measure $\nu$ on $\SX$ implies that $\int_\SX \varphi_i(\xb)\varphi_j(\xb)\, \nu(\dd\xb) = 0$ for all $i$ and $j$. When the constant $1$ belongs to the
RKHS $\SH_K$ associated with $K$,
there exists constants $\ma_i$, $i\geq 1$, such that $\sum_{i\geq 1} \ma_i\, \varphi_i(\xb)=1$ for all $\xb\in\SX$, and  $\SE_{K^2}(\nu)=0$ implies that $\int_\SX \varphi_j(\xb)\, \nu(\dd\xb) = 0$ for all $j\geq 1$. Therefore, $\SE_K(\nu)=0$, and $\nu=0$ since $K$ is ISPD. However, this argumentation cannot be combined with (\textit{i}) above to show that $\SE_{K_{|n}^2}(\xi-\mu)=0$ implies that $\xi=\mu$ since $1 \not\in \SH_{K_{|n}}$: indeed, $f(\xb_i)=0$ for any $f\in\SH_{K_{|n}}$and any $\xb_i\in\Xb_n$.
\end{description}


\vsp
\noindent{\Large\bf Appendix B: expressions of $P_{K_i^2,\mu_1}(x)$, $\SE_{K_i^2}(\mu_1)$, $\beta_{K_i}(u,v)$ and $\mg_{K_i}(u,v)$ for Mat\'ern 3/2 kernel and $\mu_1$ uniform on $[0,1]$}
\vsp

When $K_i(x,x')=K_{3/2,\mt/\sqrt{3}}(x,x')$ given in \eqref{K32}, we have \citep{GinsbourgerRSDL2014}
\bea
\SE_{K_i}(\mu_1) &=& \frac{2}{\mt^2}\, [(\mt+3)\e1^{-\mt}+2\mt-3]\,, \\
P_{K_i,\mu_1}(x) &=& S_\mt(x)+S_\mt(1-x)\,, \mbox{ with } S_\mt(x)=\frac{1}{\mt}\,[2-(2+\mt x)\e1^{-\mt x}]\,, \ x\in[0,1]\,.
\eea
Straightforward but lengthy calculation gives
\bea
\SE_{K_i^2}(\mu_1) &=& \frac{1}{4\,\mt^2}\, [(2\,\mt^2+8\,\mt+9)\e1^{-2\,\mt}+10\,\mt-9]\,, \\
P_{K_i^2,\mu_1}(x) &=& T_\mt(x)+T_\mt(1-x)\,, \mbox{ with }
T_\mt(x)=\frac{1}{4\,\mt}\,[5-(5+6\,\mt x + 2\,\mt^2 x^2)\e1^{-2\,\mt x}] \,, \ x\in[0,1]\,.
\eea
Also,
$\beta_{K_i}(u,v)= B_\mt(u,v)-C_\mt(u,v)-C_\mt(1-u,1-v)$, $u,v\in[0,1]$, with
\bea
B_\mt(u,v) &=& \frac{\e1^{-\mt|u-v|}}{6\,\mt}\,\left[15\,(1+\mt|u-v|)+6\,\mt^2|u-v|^2+\mt^3|u-v|^3\right] \,,\\
C_\mt(u,v) &=& \frac{\e1^{-\mt(u+v)}}{4\,\mt}\, \left[5+3\,\mt(u+v)+2\,\mt^2uv\right] \,,
\eea
and
$\mg_{K_i}(u,v)= G_\mt(u,1-v)+G_\mt(v,1-u)-H_\mt(u,v)-H_\mt(1-u,1-v)+I_\mt(u,v)$, $u,v\in[0,1]$, with
\bea
G_\mt(u,v) &=& \frac{\e1^{-\mt(1+u+v)}}{16\,\mt^2}\,\left\{21+\mt[9+13(u+v)]+\mt^2[6\,(u+v)+8\,uv]+4\,\mt^3 uv\right\} \,,\\
H_\mt(u,v) &=& \frac{\e1^{-\mt(u+v)}}{24\,\mt^2}\,\left\{126+96\,\mt(u+v)+24\,\mt^2(u+v)^2+3\,\mt^3(u+v)^3 \right. \\
&& \hspace{2cm} \left. +\mt^2 uv[24+6\,\mt(u+v)+2\,\mt^2(u^2+v^2)]\right\} \,, \\
I_\mt(u,v) &=& \frac{\e1^{-\mt|u-v|}}{120\,\mt^2}\,\left\{945 + 945\,\mt|u-v|+420\,\mt^2|u-v|^2+105\,\mt^3|u-v|^3+15\,\mt^4|u-v|^4 \right. \\
&& \hspace{2cm} \left. +\mt^5|u-v|^5\right\}\,.
\eea


\vsp
\noindent{\Large\bf Appendix C: random polynomials}
\vsp


Consider the family of Legendre polynomials, orthonormal for the uniform measure $\mu_1$ on $\SX_1=[0,1]$:
\bea
P_0(x) &=& 1 \\
P_1(x) &=& \sqrt{3}\,(2\,x-1) \\
P_2(x) &=& \sqrt{5}\,(6\,x^2-6\,x+1) \\
P_3(x) &=& \sqrt{7}\,(20\,x^3-30\,x^2+12\,x-1) \\
P_4(x) &=& 3\,(70\,x^4-140\,x^3+90\,x^2-20\,x+1) \\
\vdots
\eea
satisfying $\int_0^1 P_i(x)P_j(x)\, \dd x = \delta_{i,j}$ (the Kronecker delta). To each $P_i$ we associate a $\ml_i\in\mathds{R}^+$, with $\ml_0=1$ and $\ml_i>\ml_{i+1}$ for all $i$. A reasonable choice is $\ml_i=1/(i+1)^\mg$ for some $\mg>0$.
Denote by $\SSL$ a subset of $\mathds{N}^d$ containing multi-indices $\underline{\ell}=\{\ell_1,\ldots,\ell_d\}$, with each $\ell_i\in\mathds{N}$ pointing to a polynomial $P_{\ell_i}$. The multivariate polynomials we consider have the form
\be\label{Poly}
P(\xb)= \sum_{\underline{\ell}\in\SSL} \beta_{\underline{\ell}} \Psi_{\underline{\ell}}(\xb) \,,
\ee
where $\Psi_{\underline{\ell}}(\xb)=\prod_{i=1}^d P_{\ell_i}(x_i)$ and the $\beta_{\underline{\ell}}$ are independent normal variables $\SN(0,\Lambda_{\underline{\ell}})$ with $\Lambda_{\underline{\ell}}=\prod_{i=1}^d \ml_{\ell_i}$. If we only constrain the maximum degree $p$ in each variable, that is, if we consider all $\underline{\ell}$ with $\ell_i\leq p$ for all $i$, then $\SSL$ contains $(p+1)^d$ elements; if we constrain the total degree $p_T$ of $P(\xb)$, $\SSL$ has $\binom{p_T+d}{d}$ elements. In both cases, the evaluation of $f$ quickly becomes very costly when $d$, $p$ or $p_T$ increase. For that reason, we shall set a constraint on the number of elements of $\SSL$ and only retain the largest $\Lambda_{\underline{\ell}}$; that is, we use
\bea
\SSL_{N}=\{\underline{\ell}_1,\ldots,\underline{\ell}_M\in\mathds{N}^d, \mbox{ with $M$ the smaller integer } \geq N \mbox{ such that } \Lambda_{\underline{\ell}_M} < \Lambda_{\underline{\ell}_{M+1}} \}\,;
\eea
see \citet{P-RESS2019} for implementation details.

To avoid favouring too much the use of separable kernels, we apply a random linear transformation to $\xb$ before computing $f$ and set $f(\xb)=P[\Qb(\xb-\1b_d/2)+\1b_d/2]$, with
\be\label{Qrot}
\Qb = \ma \Qb_R(d) + (1-\ma) \Ib_d\,,
\ee
where $\ma\in[0,1]$, and $\Qb_R(d)$ is a random rotation matrix in the orthogonal group $\mathds{O}(d-1)$. To generate a random matrix $\Qb_R(d)$ uniformly distributed in $\mathds{O}(d-1)$ we proceed as follows, see \citet[Lemma~3.1]{DiaconisS87}. For $d=2$, we take
\bea
\Qb_R(d) = \left(
             \begin{array}{cc}
               \cos \mt & \sin \mt \\
               a\,\sin \mt & a\,\cos \mt \\
             \end{array}
           \right)\,,
\eea
with $\mt$ uniformly distributed in $[0,2\,\pi]$ and $a=\pm 1$ with probability 1/2. For larger $d$, we construct $\Qb_R(d)$ recursively as
\bea
\Qb_R(d) = \left(\Ib_d - 2\,\frac{[\eb_1-\ub(d)][\eb_1-\ub(d)]\TT}{\|\eb_1-\ub(d)\|^2} \right) \left(
                                                                                     \begin{array}{cccc}
                                                                                       1 & 0 & \cdots & 0 \\
                                                                                       0  \\
                                                                                       \vdots &  & \Qb_R(d-1) &  \\
                                                                                       0  \\
                                                                                     \end{array}
                                                                                   \right)\,,
\eea
with $\eb_1=(1,0,\ldots,0)\TT$ and $\ub(d)$ uniformly distributed on the $d$ dimensional unit sphere (for instance, we can take $\ub(d)=\vb/\|\vb\|$ with $\vb$ having the standard normal distribution $\SN(\0b_d,\Ib_d)$).
%


\section*{Acknowledgments}
This work was partly supported by project INDEX (INcremental Design of EXperiments) ANR-18-CE91-0007 of the French National Research Agency (ANR).


\bibliographystyle{apalike}

\begin{thebibliography}{}

\bibitem[Bach et~al., 2012]{BachLJO2012}
Bach, F., Lacoste-Julien, S., and Obozinski, G. (2012).
\newblock On the equivalence between herding and conditional gradient
  algorithms.
\newblock In {\em Proc.\ 29th Annual International Conference on Machine
  Learning}, pages 1355--1362.

\bibitem[Bachoc, 2013]{Bachoc2013}
Bachoc, F. (2013).
\newblock Cross validation and maximum likelihood estimations of
  hyper-parameters of {G}aussian processes with model misspecification.
\newblock {\em Comput.\ Statist.\ Data Anal.}, 66:55--69.

\bibitem[Diaconis and Shahshahani, 1987]{DiaconisS87}
Diaconis, P. and Shahshahani, M. (1987).
\newblock The subgroup algorithm for generating uniform random variables.
\newblock {\em Probability in the Engineering and Informational Sciences},
  1(1):15--32.

\bibitem[Dubrule, 1983]{Dubrule83}
Dubrule, O. (1983).
\newblock Cross validation of kriging in a unique neighborhood.
\newblock {\em Journal of the International Association for Mathematical
  Geology}, 15(6):687--699.

\bibitem[Fedorov, 1972]{Fedorov72}
Fedorov, V. (1972).
\newblock {\em Theory of Optimal Experiments}.
\newblock Academic Press, New York.

\bibitem[Gauthier and Pronzato, 2014]{GP-SIAM_2014}
Gauthier, B. and Pronzato, L. (2014).
\newblock Spectral approximation of the {IMSE} criterion for optimal designs in
  kernel-based interpolation models.
\newblock {\em SIAM/ASA J.\ Uncertainty Quantification}, 2:805--825.
\newblock {DOI} 10.1137/130928534.

\bibitem[Gauthier and Pronzato, 2016]{GP-CSSC_2016}
Gauthier, B. and Pronzato, L. (2016).
\newblock Approximation of {IMSE}-optimal designs via quadrature rules and
  spectral decomposition.
\newblock {\em Communications in Statistics -- Simulation and Computation},
  45(5):1600--1612.

\bibitem[Gauthier and Pronzato, 2017]{GP-CSDA2016}
Gauthier, B. and Pronzato, L. (2017).
\newblock Convex relaxation for {IMSE} optimal design in random field models.
\newblock {\em Computational Statistics and Data Analysis}, 113:375--394.

\bibitem[Ginsbourger et~al., 2014]{GinsbourgerRSDL2014}
Ginsbourger, D., Roustant, O., Schuhmacher, D., Durrande, N., and Lenz, N.
  (2014).
\newblock On {ANOVA} decompositions of kernels and {G}aussian random field
  paths.
\newblock {\em preprint arXiv:1409.6008}.

\bibitem[Pronzato, 2017]{P-JSFdS2017}
Pronzato, L. (2017).
\newblock Minimax and maximin space-filling designs: some properties and
  methods for construction.
\newblock {\em Journal de la Soci\'et\'e Fran\c{c}aise de Statistique},
  158(1):7--36.

\bibitem[Pronzato, 2019]{P-RESS2019}
Pronzato, L. (2019).
\newblock Sensitivity analysis via {K}arhunen-{L}o{\`e}ve expansion of a random
  field model: estimation of {S}obol' indices and experimental design.
\newblock {\em Reliability Engineering and System Safety}, 187:93--109.
\newblock hal-01545604v2.

\bibitem[Pronzato, 2021]{P2021}
Pronzato, L. (2021).
\newblock Performance analysis of greedy algorithms for minimising a maximum
  mean discrepancy.
\newblock {\em {h}al-03114891, {arXiv}:2101.07564}.

\bibitem[Pronzato and Zhigljavsky, 2020]{PZ2020-SIAM}
Pronzato, L. and Zhigljavsky, A. (2020).
\newblock Bayesian quadrature, energy minimization and space-filling design.
\newblock {\em SIAM/ASA J.\ Uncertainty Quantification}, 8(3):959--1011.

\bibitem[Santner et~al., 2003]{SantnerWN2003}
Santner, T., Williams, B., and Notz, W. (2003).
\newblock {\em The Design and Analysis of Computer Experiments}.
\newblock Springer, Heidelberg.

\bibitem[Sejdinovic et~al., 2013]{SejdinovicSGF2013}
Sejdinovic, S., Sriperumbudur, B., Gretton, A., and Fukumizu, K. (2013).
\newblock Equivalence of distance-based and {RKHS}-based statistics in
  hypothesis testing.
\newblock {\em The Annals of Statistics}, 41(5):2263--2291.

\bibitem[Sriperumbudur et~al., 2011]{SriperumbudurFL2011}
Sriperumbudur, B., Fukumizu, K., and Lanckriet, G. (2011).
\newblock Universality, characteristic kernels and {RKHS} embedding of
  measures.
\newblock {\em Journal of Machine Learning Research}, 12:2389--2410.

\bibitem[Sriperumbudur et~al., 2010]{SriperumbudurGFSL2010}
Sriperumbudur, B., Gretton, A., Fukumizu, K., Sch{\"o}lkopf, B., and Lanckriet,
  G. (2010).
\newblock Hilbert space embeddings and metrics on probability measures.
\newblock {\em Journal of Machine Learning Research}, 11:1517--1561.

\bibitem[Szab{\'o} and Sriperumbudur, 2018]{SzaboS2018}
Szab{\'o}, Z. and Sriperumbudur, B. (2018).
\newblock Characteristic and universal tensor product kernels.
\newblock {\em Journal of Machine Learning Research}, 18:1--29.

\bibitem[Wynn, 1970]{Wynn70}
Wynn, H. (1970).
\newblock The sequential generation of {$D$}-optimum experimental designs.
\newblock {\em Annals of Math. Stat.}, 41:1655--1664.

\end{thebibliography}

\end{document}